\documentstyle[12pt]{article}
\catcode`\@=11

\@addtoreset{equation}{section}
\def\theequation{\arabic{equation}}
\def\theequation{\thesection.\arabic{equation}}

\def\a{\alpha}
\def\b{\beta}

\def\e{\epsilon}

\def\NPB#1#2#3{{\it Nucl.~Phys.} {\bf{B#1}} (19#2) #3}
\def\PLB#1#2#3{{\it Phys.~Lett.} {\bf{B#1}} (19#2) #3}
\def\PRD#1#2#3{{\it Phys.~Rev.} {\bf{D#1}} (19#2) #3}
\def\PRL#1#2#3{{\it Phys.~Rev.~Lett.} {\bf{#1}} (19#2) #3}

\newskip\humongous \humongous=0pt plus 1000pt minus 1000pt
\def\caja{\mathsurround=0pt}
\def\eqalign#1{\,\vcenter{\openup1\jot \caja
        \ialign{\strut \hfil$\displaystyle{##}$&$
        \displaystyle{{}##}$\hfil\crcr#1\crcr}}\,}
\newif\ifdtup

\def\@normalsize{\@setsize\normalsize{15pt}\xiipt\@xiipt
\abovedisplayskip 14pt plus3pt minus3pt%
\belowdisplayskip \abovedisplayskip
\abovedisplayshortskip  \z@ plus3pt%
\belowdisplayshortskip  7pt plus3.5pt minus0pt}
\def\small{\@setsize\small{13.6pt}\xipt\@xipt
\abovedisplayskip 13pt plus3pt minus3pt%
\belowdisplayskip \abovedisplayskip
\abovedisplayshortskip  \z@ plus3pt%
\belowdisplayshortskip  7pt plus3.5pt minus0pt
\def\@listi{\parsep 4.5pt plus 2pt minus 1pt
            \itemsep \parsep
            \topsep 9pt plus 3pt minus 3pt}}

\def\underline#1{\relax\ifmmode\@@underline#1\else
        $\@@underline{\hbox{#1}}$\relax\fi}
\@twosidetrue
\relax

\catcode`@=12

\catcode`\@=11

\def\section{\@startsection{section}{1}{\z@}{3.5ex plus 1ex minus
   .2ex}{2.3ex plus .2ex}{\large\bf}}
\def\thesection{\arabic{section}}

\def\ps@headings{\def\@oddfoot{}\def\@evenfoot{}
\def\@oddhead{\hbox{}\hfill
        \makebox[.5\textwidth]{\raggedright\ignorespaces --\thepage{}--
        \hfill }}
\def\@evenhead{\@oddhead}
\def\subsectionmark##1{\markboth{##1}{}} }

\ps@headings

\catcode`\@=12

\relax

\def\figcap{\section*{Figure Captions\markboth
        {FIGURECAPTIONS}{FIGURECAPTIONS}}\list
        {Fig. \arabic{enumi}:\hfill}{\settowidth\labelwidth{Fig. 999:}
        \leftmargin\labelwidth
        \advance\leftmargin\labelsep\usecounter{enumi}}}
 \relax
\def\tablecap{\section*{Table Captions\markboth
        {TABLECAPTIONS}{TABLECAPTIONS}}\list
        {Table \arabic{enumi}:\hfill}{\settowidth\labelwidth{Table 999:}
        \leftmargin\labelwidth
        \advance\leftmargin\labelsep\usecounter{enumi}}}
 \relax
\def\reflist{\section*{References\markboth
        {REFLIST}{REFLIST}}\list
        {[\arabic{enumi}]\hfill}{\settowidth\labelwidth{[999]}
        \leftmargin\labelwidth
        \advance\leftmargin\labelsep\usecounter{enumi}}}
 \relax

\catcode`\@=11

\def\marginnote#1{}
\newcount\hour
\newcount\minute
\newtoks\amorpm
\hour=\time\divide\hour by60
\minute=\time{\multiply\hour by60 \global\advance\minute by-
\hour}
\edef\standardtime{{\ifnum\hour<12 \global\amorpm={am}%
    \else\global\amorpm={pm}\advance\hour by-12 \fi
    \ifnum\hour=0 \hour=12 \fi
    \number\hour:\ifnum\minute<100\fi\number\minute\the\amorpm}}
\edef\militarytime{\number\hour:\ifnum\minute<100\fi\number\minute}
\def\draftlabel#1{{\@bsphack\if@filesw {\let\thepage\relax
  \xdef\@gtempa{\write\@auxout{\string
    \newlabel{#1}{{\@currentlabel}{\thepage}}}}}\@gtempa
    \if@nobreak \ifvmode\nobreak\fi\fi\fi\@esphack}
     \gdef\@eqnlabel{#1}}
\def\@eqnlabel{}
\def\@vacuum{}
\def\draftmarginnote#1{\marginpar{\raggedright\scriptsize\tt#1}}
\def\draft{\oddsidemargin -.5truein
        \def\@oddfoot{\sl preliminary draft \hfil
        \rm\thepage\hfil\sl\today\quad\militarytime}
        \let\@evenfoot\@oddfoot \overfullrule 3pt
        \let\label=\draftlabel
        \let\marginnote=\draftmarginnote
   
\def\@eqnnum{(\theequation)\rlap{\kern\marginparsep\tt\@eqnlabel}%
\global\let\@eqnlabel\@vacuum}  }
\def\preprint{\twocolumn\sloppy\flushbottom\parindent 1em
        \leftmargini 2em\leftmarginv .5em\leftmarginvi .5em
        \oddsidemargin -.5in    \evensidemargin -.5in
        \columnsep 15mm \footheight 0pt
        \textwidth 250mmin      \topmargin  -.4in
        \headheight 12pt \topskip .4in
        \textheight 175mm
        \footskip 0pt
        
\def\@oddhead{\thepage\hfil\addtocounter{page}{1}\thepage}
        \let\@evenhead\@oddhead \def\@oddfoot{} \def\@evenfoot{}  }
\def\titlepage{\@restonecolfalse\if@twocolumn\@restonecoltrue\onecolumn
     \else \newpage \fi \thispagestyle{empty}\c@page\z@
        \def\thefootnote{\fnsymbol{footnote}} }
\def\endtitlepage{\if@restonecol\twocolumn \else  \fi
        \def\thefootnote{\arabic{footnote}}
        \setcounter{footnote}{0}}  
\catcode`@=12
\relax

\def\ps@headings{\def\@oddfoot{}\def\@evenfoot{}
\def\@oddhead{\hbox{}\hfill
        \makebox[.5\textwidth]{\raggedright\ignorespaces --\thepage{}--
        \hfill }}
\def\@evenhead{\@oddhead}
\def\subsectionmark##1{\markboth{##1}{}} }

\ps@headings

\relax

\def\firstpage#1#2#3#4#5#6{
\begin{document}
\begin{titlepage}
\nopagebreak
\title{\begin{flushright}
        \vspace*{-1.8in}
        {\normalsize LPT-ORSAY 99/88}\\[-9mm]
        {\normalsize LPTM-99/56 }\\[-9mm]
\end{flushright}
\vfill {#3}}
\author{\large #4 \\[1.0cm] #5}
\maketitle
\vskip -10mm     
\nopagebreak 
\begin{abstract} {\noindent #6}
\end{abstract}
\vfill
\begin{flushleft}
\end{flushleft}
\thispagestyle{empty}
\end{titlepage}}

\def\simlt{\stackrel{<}{{}_\sim}}
\def\simgt{\stackrel{>}{{}_\sim}}
\newcommand{\dal}{\raisebox{0.085cm} {\fbox{\rule{0cm}{0.07cm}\,}}}
\newcommand{\dt}{\partial_{\langle T\rangle}}
\newcommand{\dtbar}{\partial_{\langle\overline{T}\rangle}}
\newcommand{\al}{\alpha^{\prime}}
\newcommand{\mst}{M_{\scriptscriptstyle \!S}}
\newcommand{\mpl}{M_{\scriptscriptstyle \!P}}
\newcommand{\dv}{\int{\rm d}^4x\sqrt{g}}
\newcommand{\lv}{\left\langle}
\newcommand{\rv}{\right\rangle}
\newcommand{\ph}{\varphi}
\newcommand{\abar}{\overline{a}}
\newcommand{\sbar}{\,\overline{\! S}}
\newcommand{\xbar}{\,\overline{\! X}}
\newcommand{\fbar}{\,\overline{\! F}}
\newcommand{\zbar}{\overline{z}}
\newcommand{\dbar}{\,\overline{\!\partial}}
\newcommand{\tbar}{\overline{T}}
\newcommand{\taubar}{\overline{\tau}}
\newcommand{\ubar}{\overline{U}}
\newcommand{\ybar}{\overline{Y}}
\newcommand{\phb}{\overline{\varphi}}
\newcommand{\cm}{Commun.\ Math.\ Phys.~}
\newcommand{\prl}{Phys.\ Rev.\ Lett.~}
\newcommand{\pr}{Phys.\ Rev.\ D~}
\newcommand{\pl}{Phys.\ Lett.\ B~}
\newcommand{\ibar}{\overline{\imath}}
\newcommand{\jbar}{\overline{\jmath}}
\newcommand{\np}{Nucl.\ Phys.\ B~}
\newcommand{\F}{{\cal F}}
\renewcommand{\L}{{\cal L}}
\newcommand{\A}{{\cal A}}
\newcommand{\be}{\begin{equation}}
\newcommand{\ee}{\end{equation}}
\newcommand{\ba}{\begin{eqnarray}}
\newcommand{\ea}{\end{eqnarray}}
\newcommand{\dslash}{{\not\!\partial}}
\newcommand{\gsi}{\,\raisebox{-0.13cm}{$\stackrel{\textstyle >}{\textstyle\sim}$}\,}
\newcommand{\lsi}{\,\raisebox{-0.13cm}{$\stackrel{\textstyle <}{\textstyle\sim}$}\,}

\firstpage{3118}{IC/95/34} 
{\large\bf String theory predictions for future accelerators} 
{E. Dudas$^{\,a}$\footnote{e.mail: dudas@qcd.th.u-psud.fr} and 
J. Mourad$^{\,b}$\footnote{e.mail: mourad@qcd.th.u-psud.fr}}
{\normalsize\sl
$^a$ LPT\footnote{{\small Unit{\'e} mixte de recherche du CNRS (UMR 8627).}}, B{\^a}t. 210, Univ. de Paris-Sud, F-91405 Orsay,
France \\[-3mm]
\normalsize\sl$^b$ LPTM, Site  Neuville III,
Univ. de Cergy-Pontoise, Neuville sur Oise\\[-3mm]
\normalsize\sl F-95031 Cergy-Pontoise, France} 
{
We consider, in a string theory framework, physical processes of phenomenological
interest in models with a low string scale. The amplitudes we study
involve tree-level virtual gravitational exchange, divergent in
a field-theoretical treatment, and massive gravitons emission, which
are the main signatures of this class of models.  
First, we discuss the regularization of summations appearing in
virtual gravitational (closed string) Kaluza-Klein exchanges in Type I 
strings. We argue that a convenient manifestly ultraviolet convergent
low energy limit of type I string theory is given by an effective field theory 
with an arbitrary cutoff $\Lambda$ in the closed (gravitational)
channel and a related cutoff $M_s^2/\Lambda$ in the open (Yang-Mills)
channel. We find the leading string corrections to the field theory
results. Second, we calculate exactly string tree-level three and four-
point amplitudes with gauge bosons and  one massive graviton and examine 
string deviations from the field-theory result.
}
\section{Introduction and Summary of the results.}

Shortly after the birth of string theory as a theory of hadronic
interactions with a mass scale of the order of nucleon masses, it was 
realized that string theory is
actually the natural framework to quantize gravity
\cite{gsw}. For a long time, the phenomenologically most interesting theories
were considered to be the heterotic strings, where the string scale is
of the order of the Planck scale. This rendered string theory
predictions not directly accessible to current or future
accelerators, with the notable exception of some peculiar models
\cite{antoniadis}. Recent progress in the understanding of string dualities and D-branes
\cite{polchinski} led to other string constructions \cite{witten}, where
the string scale can have values directly accessible 
in future accelerators.

Consequently, a lot of efforts were made in order to understand the main
features of low-scale string theories, from the point of view
of possible existence of submilimeter dimensions which can provide
testable deviations from the Newton law \cite{ADD}, gauge coupling 
unification \cite{DDG} and corresponding string embedding \cite{AADD}. The main
interest of these theories comes from their possible testability at the
future colliders, through the direct production or indirect (virtual)
effects of Kaluza-Klein states \cite{GRW} in various cross-sections.  
This paper is devoted to the (Type I) string computations of the
relevant amplitudes. For the convenience of the reader we provide
in the following a brief summary of our results.

A subtle issue concerning the virtual effects of gravitational
Kaluza-Klein particles is that for a number of compact dimensions
$d \ge 2$ the corresponding field theory summations diverge in the ultraviolet (UV).
Indeed, let us consider a 
four-fermion interaction of particles stuck on a D3 brane mediated by
Kaluza-Klein gravitational excitations orthogonal to it. Then the 
amplitude of the process, depicted in Figure 1, reads
\be
A \ = \ {1 \over M_P
^2} \sum_{m_i} {1 \over -s +
{m_1^2 + \cdots m_{d}^2 \over R_{\perp}^2}} \ , \label{I1}
\ee
where for simplicity we considered equal radii denoted by $R_{\perp}$ 
and $s=-(p_1+p_2)^2$ is the squared center of mass energy\footnote{Within our conventions
$s$ is negative in Euclidean space.}.
The summation clearly diverges for $d \ge 2$. 
\begin{figure}
\vspace{4 cm}
\special{hscale=60 vscale=60 voffset=0 hoffset=120
psfile=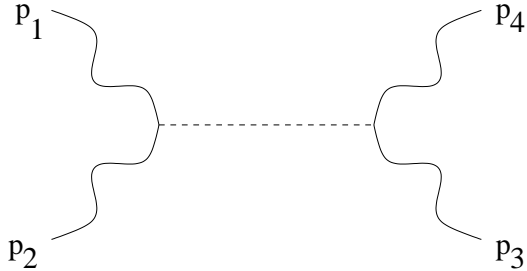}
\caption{Tree-level gravitational virtual exchange.}
\end{figure}

The traditional attitude to
adopt in this case is to cut the sums for masses heavier than a cutoff
$\Lambda >> R_{\perp}^{-1}$, of the order of the fundamental scale 
$M_s$ in the string theory \cite{GRW} . This can be implemented in a
proper-time representation of the amplitude
\be
A \ = \ {1 \over M_P^2} \sum_{m_i} \int_{1 / \Lambda^2}^{\infty} dl \ e^{-l
(-s + {m_1^2 + \cdots m_{d }^2 \over R_{\perp}^2})}
\ = \  {1 \over M_P^2} \int_{1 / \Lambda^2}^{\infty} dl \ e^{sl}
\ \theta_3^{d } (0,{il \over \pi R_{\perp}^2}) \ , \label{I2}
\ee 
where $\theta_3 (0,\tau)=\sum_k exp(i \pi k^2 \tau)$ is one of the Jacobi
functions. We shall be interested in the following in the region of
the parameter space
$-R_{\perp}^2 s >>1$, $R_{\perp} \Lambda >>1$ and  
$-s << \Lambda^2$ in which the available energy is smaller (but not far
away) from  the UV cutoff $\Lambda$ but much bigger than the (inverse)
compact radius $R_{\perp}^{-1}$, of submilimeter size. In this case, the
amplitude can be evaluated to give
\be
A \ = \ {\pi^{d} R_{\perp}^{d} \over M_P^2} \int_{1 /
\Lambda^2}^{\infty} {dl \over l^{d \over 2}} \ e^{sl} \ 
\theta_3^{d } (0,{i \pi R_{\perp}^2 \over l}) \ \simeq \ 
{2 \pi^{d \over 2} \over d-2} {R_{\perp}^{d} \Lambda^{d-2} \over
M_P^2} \ = \ {4 \pi^{d \over 2} \over d-2} \alpha_{YM}^2 {\Lambda^{d-2} \over
M_s^{d+2}} \ , \label{I3}
\ee
where in the last step we used the relation
$M_P^2=(2/\alpha_{YM}^2)R_{\perp}^{d} M_s^{2+d}$, valid for Type I strings,
where $\alpha_G=g_{YM}^2/(4 \pi)$ and $g_{YM}$ 
is the Yang-Mills coupling on our brane.
The high sensitivity of the result on the cutoff asks for a more precise
computation in a full Type I string context.
This is one of the aims of this paper. In what follows we present 
qualitatively the results  which we derive in Section 3.

The computation in the following is done for the $SO(32)$ Type I 10D 
superstring compactified down to 4D on a six-dimensional torus. However,
as we shall argue later on, the result holds for a large class of
orbifolds, including ${\cal N}=2$ and ${\cal N}=1$ 
supersymmetric vacua \cite{sagnotti}.
The Type I string diagram which contains in the low-energy limit the 
gravitational exchange mentioned above is the nonplanar cylinder diagram
depicted in Fig.2, in which for simplicity we prefer to put gauge
bosons instead of fermions in the external lines. This diagram has a
twofold dual interpretation \cite{dua}
a) tree-level exchange of closed-string
states, if the time is chosen to run horizontally (see Fig. 3)
b) one-loop diagram of open strings, if the time runs vertically in 
the diagram (see Fig. 4). In the two dual representations, the
nonplanar amplitude reads symbolically
\ba
A &=& \sum_n \int_0^{\infty} \ dl \sum_{n_i} \ A_2 (l,n_1 \cdots n_d,n)
\nonumber \\
&=&   \sum_{k_1 \cdots k_4} \int_0^{\infty} \ d\tau_2 \tau_2^{d/2-2} 
\sum_{m_i} \ A_1 (\tau_2,m_1 \cdots m_d,k_1 \cdots k_4) \ , \label{I4} 
\ea
where $l$ denotes the cylinder parameter in the tree-level channel and
$\tau_2=1/l$ is the one-loop open string parameter. In the first
representation, the amplitude is interpreted as tree-level exchange of
closed-string particules of mass $(n_1^2+ \cdots n_d^2) R^2M_s^4 + n
M_s^2$, where $n_1 \cdots n_d$ are winding quantum numbers and $n$ is
the string oscillator number. In particular the $n=0$ term reproduces
the field-theory result (\ref{I1}) and therefore the full expression
(\ref{I4}) is its string regularization.
In the second representation, the
amplitude is interpreted as a sum of box diagrams with particles of
masses $(m_1^2+\cdots m_d^2)/R^2+k_i M_s^2$ ($i=1 \cdots 4$) running in 
the four propagators of the diagram.

\begin{figure}
\vspace{4 cm}
\special{hscale=60 vscale=60 voffset=0 hoffset=120
psfile=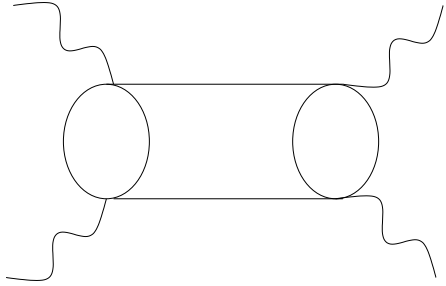}
\caption{The string nonplanar amplitude.}
\end{figure}
\begin{figure}
\vspace{4 cm}
\special{hscale=60 vscale=60 voffset=0 hoffset=100
psfile=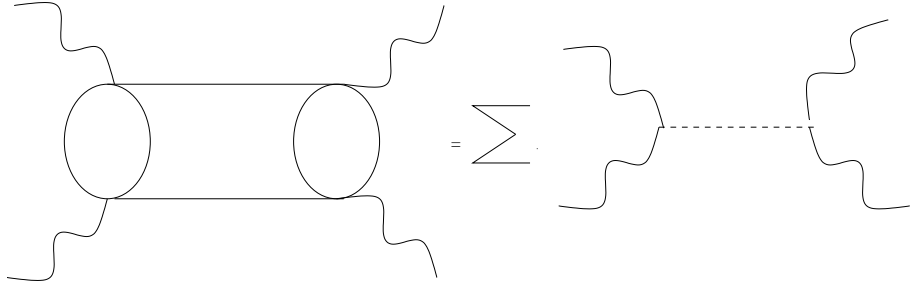}
\caption{The closed channel representation of the amplitude.}
\end{figure}
\begin{figure}
\vspace{4 cm}
\special{hscale=60 vscale=60 voffset=-5 hoffset=100
psfile=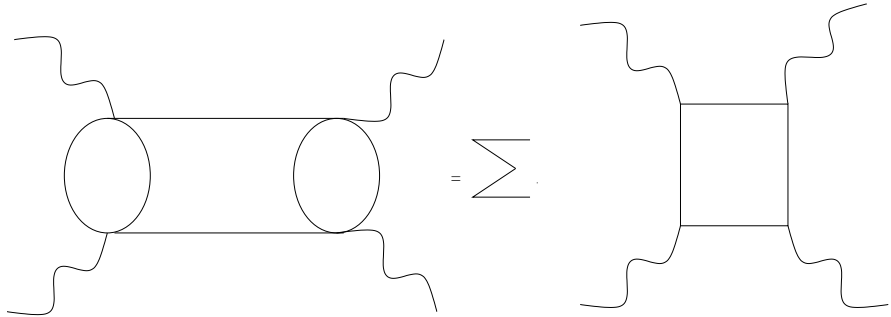}
\caption{The open channel representation of the amplitude.}
\end{figure}

The UV limit ($l \rightarrow 0$) of the gravitational 
tree-level diagram is related to the IR limit ($\tau_2
\rightarrow \infty$) of the box diagram.
In particular, in four dimensions when an IR regulator
$\mu$ is introduced in
the box diagram, the divergence in the Kaluza-Klein (KK) summation in the 
gravitational-exchange diagram cancels out. 
The final result for the
nonplanar cylinder amplitude in the low energy limit $E/M_s <<1$ ($E$ is
a typical energy scale), which is one of the main results of this paper
to be discussed in Section 3.5,  is in four-dimensions (D=4)
\ba
A &=& -{1 \over \pi M_P^2 s} + {2 g_{YM}^4 \over \pi^2 } \ [ \ {1 \over st} 
\ln {-s \over
4 \mu^2} \ln {-t \over 4 \mu^2}+ {\rm perms.}]  \nonumber \\
&-& {g_{YM}^4 \over 3 M_s^4} 
\ [ \ \ln {s \over t}  \ln {s t \over \mu^4} + \ln {s \over u} \ln {s u
\over \mu^4} ]+ \cdots \ , \label{I5}
\ea
where ${\rm perms.}$ denotes two additional contributions coming from the
permutations of $s,t$, $u$ and $\cdots$ denote terms of higher order
in the low energy expansion. Notice in (\ref{I5}) the absence of the
contact term (\ref{I3}) in the string result, which is replaced by the
leading string correction, given by the second line in (\ref{I5}).
The string correction in (\ref{I5}) is indeed of the same order of
magnitude as (\ref{I3}) for $\Lambda \sim M_s$, however it has an
explicit energy dependence coming from the logarithmic terms.
 
In order to find the appropriate interpretation of (\ref{I5}) in terms
of field-theory diagrams, it is  convenient to separate the integration region
in (\ref{I4}) into two parts, by introducing an arbitrary parameter $l_0$ and writing
\be
A = \sum_n \int_{l_0}^{\infty} \ dl \sum_{n_i} \ A_2 \ +
 \sum_{k_1 \cdots k_4} \int_{1/l_0}^{\infty} \ d\tau_2 \tau_2^{d/2-2} 
\sum_{m_i} \ A_1 \ . \label{I6}
\ee  
This has the effect of fixing an UV cutoff $\Lambda=M_s/\sqrt{l_0}$ in the tree-level
exchange diagram, similar to the
one introduced in (\ref{I2}), (\ref{I3}),
and simultaneously of a related UV cutoff $\Lambda'=M_s \sqrt{l_0}=M_s^2/\Lambda$ in
the one-loop box diagram described here by $A_1$. 
This "mixed" repesentation of the non planar amplitude
is depicted in Figure 5. 
By computing the low-energy limit of $A_1$ and $A_2$ we find in D=4
\ba  
A_1 &=& {2 g_{YM}^4 \over \pi^2} \ [ \ {1 \over st} \ln {-s \over
4 \mu^2} \ln {-t \over 4 \mu^2}+ {\rm perms.}] - 
{g_{YM}^4 \over 3 M_s^4} 
\ [ \ \ln {s \over t}  \ln {s t \over \mu^4} + \ln {s \over u} \ln {s u
\over \mu^4} + {6 \over l_0^2}] + \cdots \nonumber \\
A_2 &=&  -{1 \over \pi M_P^2 s}+ {2g_{YM}^4 \over M_s^4}
\ [ \ {1 \over l_0^2} + \cdots + O({s^2 \over M_s^4})+ \cdots] \  . \label{I7}
\ea
The $g_{YM}^4$ terms in $A_1$ describe a box diagram with four light
particles (of mass $\mu$) circulating in the loop, while the
$g_{YM}^4/M_s^4$ terms are the first string corrections coming from box
diagrams with one massive particle (of mass $M_s$) and three light
particles of mass $\mu$ in the loop. It contains also the $l_0$ dependent
part of the box diagram with four light particles in the loop. 
The $1/M_s^4l_0^2$ term in $A_2$ can be written as $\Lambda^4 /
M_s^8$ and reproduces therefore the field theory computation (\ref{I3})
in the case $d=6$. However, as expected, a similar term with opposite
sign appears in $A_1$ and the $l_0$ dependent terms cancel. 
In $A_2$ the first dots contain $l_0$ dependent terms which cancel with 
higher-order contributions in $A_1$ and the second dots denote higher-order 
contributions, while the  $O({s^2 / M_s^4})$ term is $l_0$ independent
and is actually the first correction to the tree-level graviton exchange.
We emphasize, however, that the only physically meaningful 
amplitude is the full expression (\ref{I5}) and the leading
string correction is therefore the second line of (\ref{I5}), coming
from box diagrams $A_1$ with one massive particle in the loop.
 
\begin{figure}
\vspace{7 cm}
\special{hscale=60 vscale=60 voffset=0 hoffset=60
psfile=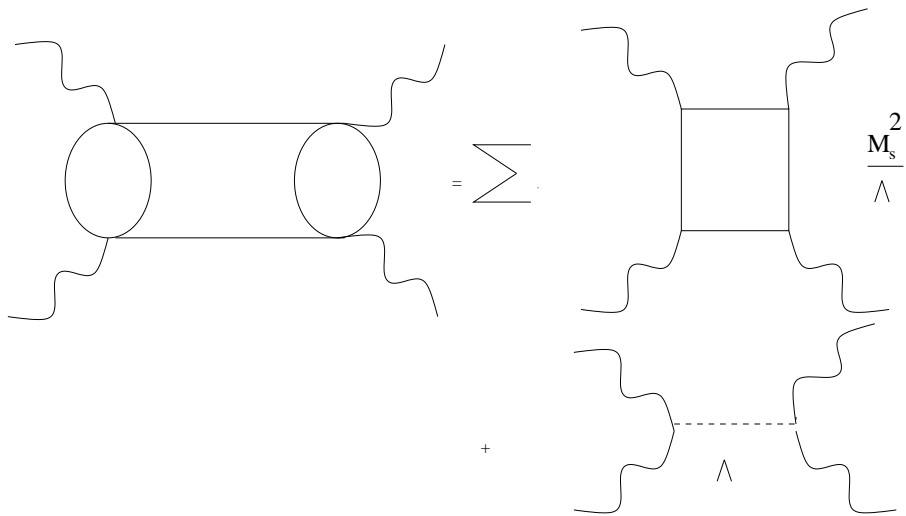}
\caption{The mixed representation of the amplitude
where the tree diagrams have a cutoff $\Lambda$ and the box
diagrams a cutoff $M_s^2/\Lambda$.}
\end{figure}

Strictly speaking, the result described above is valid for the toroidal
compactification of the $SO(32)$ 10D Type I string. For a general ${\cal
N}=1$ supersymmetric 4D Type I vacuum the amplitude $A$ has
contributions from sectors with various numbers of supersymmetries
\be
A= A^{{\cal N}=4}+ A^{{\cal N}=2}+ A^{{\cal N}=1} \ , \label{I8}
\ee
where the ${\cal N}=4$ sector contains the six-dimensional compact KK
summations, ${\cal N}=2$ sectors contain two-dimensional compact KK
summations and ${\cal N}=1$ sectors contain no KK summations. From the
tree-level ($A_2$) viewpoint, the ${\cal N}=2$ sectors give logarithmic
divergences which correspond in the one-loop box ($A_1$) picture to additional
infrared divergences associated to wave-functions or vertex corrections,
which were absent (by nonrenormalization theorems) for the ${\cal N}=4$ theory. 
Similarly, ${\cal N}=1$ sectors give no KK divergences. As the important 
(power-type) divergences come from the gravitational ${\cal N}=4$
sector, the toroidally compactified Type I superstring
contains therefore the relevant information for our purposes.
Moreover, even if we place ourselseves in the context of Type I
superstring, the formalism we use can be easily adapted to a Type II
string context and the associated D-branes. This can be done by exchanging some of the
Neumann boundary conditions in the compactified Type I string with the
appropriate Dirichlet ones for the D-branes \cite{HK,gm}. The basic results and
conclusions of our paper can be easily seen to be unchanged.  

The second aim of our paper is to calculate the tree-level string
amplitudes with two and three gauge bosons and one winding graviton
emission. For theories with low string scale and (sub)millimeter
dimensions, this type of processes is one of the best signals for future
accelerators and was computed in field theory in \cite{GRW}.
A full string formula is needed, however, for energies close to the
string scale where string effects are important.
We start by computing the two gauge bosons -- one winding (KK mode
${\bf m}$ after T-dualities) graviton
amplitude. The resulting expression has poles and zeroes for discrete
values of energies, to be explained in Section 4. 
We then compute the technically
more difficult and phenomenologically more interesting amplitude for
three gauge bosons and  one massive graviton. We study the deviations from the 
field-theory result and show that they are of order ${\bf m}^4/(R_{\perp}M_s)^4$.
The full amplitude has an interesting structure of poles and
zeroes and allows, as explained in Section 4, to define an 
off-shell form factor. 
By combining the results of Sections 3 and 4, the effective vertex of two gauge bosons 
(one of which can be off-shell) of momenta $p_1,p_2$ 
and an off-shell graviton of momentum $p$ (see Figure 6) can be written as
\be
{1 \over M_P \sqrt{\pi}} \ 2^{-p^2 \over M_s^2}{\Gamma
(-p^2/2M_s^2+1/2) \over
\Gamma (-p_1p_2/2M_s^2+1)} \ . \label{I9}
\ee
From this we can deduce a form factor characterizing heavy graviton emission
($p^2 >> M_s^2$)
\be
g(p^2) \sim 2 \sqrt{2M_s^2 \over \pi p^2} 
(\tan{\pi p^2 \over M_s^2}) e^{-{p^2 \over M_s^2} \ln 2} \ , \label{I10}            
\ee
where for an on-shell graviton $p^2$ is equal to the KK graviton 
mass $p^2=m^2/R_{\perp}^2$.

The plan of the paper is as follows. In Section 2 we review the mass
scales and coupling constants in Type I string compactified on torii.
Section 3 is devoted to the study of the virtual gravitational exchange.
As explained above, this amounts in a Type I context to a one-loop
nonplanar cylinder diagram described in Section 3.1. Sections 3.2 and
3.3 give two dual field-theoretical interpretations of the amplitude as
one-loop box diagrams with open modes circulating in the loop and 
tree-level (winding) gravitational exchange, respectively. A
representation of the amplitude suitable for the low-energy manifestly
UV convergent expansion is provided in Section 3.4 and applied to the
compactified Type I string in Section 3.5, where the first string
corrections to the field theory amplitude are computed. In Section 4
we consider the tree-level (disk) one-graviton emission amplitudes with
two gauge bosons in Section 4.1 and three gauge bosons in Section 4.2. 
Finally, Appendix A contains definitions and some properties of
Jacobi theta functions, Appendix B calculations of 4D box diagrams
and Appendix C some details on the disk tree-level amplitudes of
Section 4.2.

\section{Coupling constants}

Consider the type I superstring compactified 
to $D=10-d$ dimensions on a torus $T^d$
with (equal for simplicity) radii $R$. 
The D-dimensional Planck mass 
and Yang-Mills
coupling constant are given in terms of the string scale $M_s$
and the string coupling constant $g_s$  by
\begin{equation}
M_{P}^{8-d}={{R^d M_s^8}\over{g_s^2}},\quad
g^{-2}_{YM}={{R^d M_s^6}\over{g_s}} \ .\label{coup}
\end{equation}
Eliminating the radius $R$ in the above two relations for $D=4$ we get
\begin{equation}
\lambda \equiv{{M_P^2 \over M_s^2}}={1 \over{g_s g_{YM}^2}} \ ,
\end{equation}
which shows that the ratio $M_P/M_s$ can be very large
if the string coupling constant $g_s$ is very small.
The radius $R$ can be determined in terms of $M_s$
and the string and Yang-Mills coupling constants as
\begin{equation}
(RM_s)^6 ={{g_s} \over{g_{YM}^2}}={{1}\over{\lambda g_{YM}^4}}.
\end{equation}
So if the string scale is much lower than the 
four dimensional Planck scale, that is $\lambda\gg 1$, then
the radius $R$ is very small compared to the string length
$RM_s\ll 1$.

The equivalent T-dual description is given by a type II theory
on $T'{}^{6}$ with 32 D3-branes and 64 orientifold planes.
The radius of $T'$ is given by $R_{\perp}M_s=(RM_s)^{-1}$,
so it is very large compared to the string length. The 
T-dual string coupling constant is given by
\begin{equation}
g_s'^2=g_s^2 \left({{R}_{\perp} \over{R}}\right)^6
= g_{YM}^4 \ .
\end{equation}
Let $E$ be the order of magnitude 
energy in a physical process. We shall mainly be interested
in the low energy regime where $ E / M_s \ll 1$.
Moreover we shall suppose that $\lambda^{-1}\ll E /M_s$,
which is compatible with a low string scale.
The low energy limit of type I superstrings
was considered by Green, Schwarz and Brink \cite{gsb} in the regime 
$E / M_s \ll\lambda^{-1}$ with $\lambda$ fixed, which corresponds
to the gravitational decoupling limit $M_P \rightarrow \infty$.
It was shown there that this limit is given by the ${\cal N}=4$ super
Yang-Mills finite theory in four-dimensions. 
In the following, Section 3.1, we look for the leading stringy and KK
corrections to the four-point amplitude described in the Introduction 
for values of parameters mentioned above and by keeping a finite value 
for $M_P$.

\section{Virtual gravitational exchange amplitude}

\subsection{One loop type I amplitudes}

The one loop type I amplitudes for the scattering 
of four external massless gauge
bosons of momenta $p_i$, polarisation $\epsilon_i^\mu$,
and Chan-Paton factors $\lambda_i$ are of the form
\begin{equation}
{\cal A}_{\alpha}(p,\epsilon,\lambda)=
\delta(\sum p_i)G_{\alpha}
K_{\mu_1,\dots\mu_4}\epsilon_1^{\mu_1}\dots\epsilon_4^{\mu_4}
A_{\alpha}(s,t,u) \ ,
\end{equation}
where the index $\alpha=1,2,3$ labels the three diagrams that
contribute to the one loop level, the planar cylinder, nonplanar
cylinder we are interested in and the M\"obius amplitude.
For the non planar cylinder with two vertex operators at each
boundary, the corresponding group theory factor $G_{\alpha}$ is
\begin{equation}
G=tr(\lambda_1\lambda_2)tr(\lambda_3\lambda_4) \ .
\end{equation}
The kinematical factor $K$ is a polynomial in the 
external momenta and is given by
\begin{eqnarray}
&&\!\!\!K_{\mu_1\dots\mu_4}\!=\!-(st\eta_{13}\eta_{24}+su\eta_{14}\eta_{23}
+tu\eta_{12}\eta_{34}) +s(p_1^4p_3^2\eta_{24}+
p_2^3p_4^1\eta_{13}+p_1^3p_4^2\eta_{23}+p_2^4p_3^1\eta_{14})
\nonumber\\
&\!\!+\!\!&t(p_2^1p_4^3\eta_{13}\!+\!p_3^4p_1^2\eta_{24}
\!+\!p_2^4p_1^3\eta_{34}\!+\!p_3^1p_4^2\eta_{12}) \!+\!
u(p_1^2p_4^3\eta_{23}\!+\!p_3^4p_2^1 \eta_{14}\!+\!
p_1^4p_2^3\eta_{34}\!+\!p_3^2p_4^1\eta_{12}) \ ,
\end{eqnarray}
where the upper index labels the external particles
and the lower index $i$ is an abreviation of the Lorentz
index $\mu_i$. We have also used the 
Mandelstam variables
\begin{equation}
s=-(p_1+p_2)^2,\quad t=-(p_1+p_4)^2,\quad u=-(p_1+p_3)^2 \ , \label{ma}
\end{equation}
that verify $s+t+u=0$.

The amplitudes $A_\alpha$ can be written as integrals
over the modular parameter $\tau_2$ of the corresponding 
surface and the positions $w_i$ of the vertex operators
\begin{equation}
A_{\alpha}= {{g_s^2}\over{M_s^{10}}} \int_{0}^{\infty}{{d\tau_2}\over{\tau_2^2}}
\int_{R_{\alpha}}dw \ \prod_{i>j} \exp\left({p_i.p_j \over M_s^2}
G(w_i-w_j)\right),\label{am}
\end{equation}
where $G_{\alpha}$ is the Green function on the corresponding
surface. It can be expressed with the aid of the 
Green function on the torus
\begin{equation}
G(z,\tau)=-\ln\left|{{\vartheta_1(z,\tau)}\over{\vartheta'_1 (0,\tau)}}
\right|^2
+{{2\pi}\over{\tau_2}}( Im{(z)})^2 \ .
\end{equation}
The definitions and some useful properties of Jacobi modular functions
$\theta_i(z,\tau)$ are given in Appendix A.
From (\ref{f8}) we get 
\begin{equation}
G\left(z/(c\tau+d),(a\tau+b)/(c\tau+d)\right)=
G(z,\tau)+\ln|c\tau+d| \ , \label{mod1}
\end{equation}
where $a,b,c$ and $d$ are elements of an SL(2,Z) matrix.

The Green function on the cylinder 
is conveniently obtained from that of its covering torus.
Let the torus with modular parameter $\tau=\tau_1+i\tau_2$ be
parametrised by the complex coordinate
$w$ with $w=w+1=w+\tau$ and let $w=x+\tau\nu$,
$x$ and $\nu$ being two real 1-periodic coordinates.
Then the cylinder is obtained by setting $\tau_1=0$ and 
orbifolding with $w=-\bar w$, the  two boundaries being 
at $x=Re(w)=0,1/2$. The parameter $\tau_2$ represents the 
length of the circles at the boundaries. 
In the amplitudes (\ref{am}), the region of integration over the positions $w$
is given by $\nu_i<\nu_{i+1}$ whenever the two 
vertex operators are on the same boundary, the value of $\nu_4$
being fixed to 1 and the coordinate $x$ is fixed  
for the cylinder at $0$ or $1/2$.
It will be convenient to use the notations:
\begin{eqnarray}
\Psi(\nu,\tau_2)&=&\exp\left({{1}\over{2}}
G(i\tau_2\nu,i\tau_2)\right),\quad
\Psi^{T}(\nu,\tau_2)=\exp\left({{1}\over{2}}G({{1}\over{2}}
+i\tau_2\nu,i\tau_2)\right) \ . \label{o1}
\end{eqnarray}
With these notations the amplitude $A$ can be cast 
in the form (where $\psi_{12}$ stands for
$\Psi(\nu_2-\nu_1,i\tau_2)$ and so on) \cite{gs}
\be
A = {{g_s^2}\over{M_s^{10}}} \int_{0}^{\infty}{{d\tau_2}\over{\tau_2^2}}\ 
\int_{0}^{1}d\nu_2\int_{0}^{\nu_2}d\nu_1
\int_{0}^{1}d\nu_3\ 
\left({{\Psi^T_{13}\Psi^T_{24}}
\over{\Psi_{12}\Psi_{34}}}\right)^{s/M_s^2}
\left({{\Psi^T_{13}\Psi^T_{24}}
\over{\Psi^T_{14}\Psi^T_{23}}}\right)^{t/M_s^2} F_{6}(\tau_2,R_i) \ ,
\label{o4}
\ee
where
\be
F_{d}(\tau_2,R_i)= {M_s^{d} (2\tau_2)^{d/2} \over (R M_s)^{2d}}
\sum_{m_i}\exp{\left(-2\pi\tau_2\alpha'\sum_i 
{{m_i^2}\over{R^2}}\right)} ={M_s^{d} (2\tau_2)^{d/2} 
\over{(R M_s)^{2d}}} 
\vartheta_3^d \left(0,{{2i\tau_2}\over{(RM_s)^2}}\right) \ 
\ee
is a factor coming from the toroidal compactification on torii with
radii (taken equal for simplicity) $R$.

The transformation of the torus Green function
under the modular group suggests the possibility of using
other modular parameters than $\tau_2$. Defining $l=1 / {\tau_2}$, 
the transformation (\ref{mod1}) gives
\begin{equation}
G(z,i\tau_2)=G(z/(i\tau_2),i/\tau_2)-\ln{(\tau_2)} \ ,
\end{equation}
which implies that
\begin{eqnarray}
\Psi(\nu,\tau_2)&=&{{\sqrt{l}}}
\exp{\left({{1}\over{2}}G(\nu,il)\right)}
\equiv \tilde\Psi(\nu,l) \ ,\nonumber \\
\Psi^T(\nu,\tau_2)&=&\sqrt{l}
\exp{\left({{1}\over{2}}G(\nu-i{{l}\over{2}},il)\right)}
\equiv\tilde\Psi^T(\nu,l) \ . \label{psi}
\end{eqnarray}
With the new modular parameter $l$
the amplitude can be cast in the following form
\be
A = {{g_s^2}\over{M_s^{10}}} \int_{0}^{\infty}dl\ 
\int_{0}^{1}d\nu_2\int_{0}^{\nu_2}d\nu_1
\int_{0}^{1}d\nu_3\ 
\left({{\tilde\Psi^T_{13}\tilde\Psi^T_{24}}
\over{\tilde\Psi_{12}\tilde\Psi_{34}}}\right)^{s/M_s^2}
\left({{\tilde\Psi^T_{13}\tilde\Psi^T_{24}}
\over{\tilde\Psi^T_{14}\tilde\Psi^T_{23}}}\right)^{t/M_s^2} F_6 
\ . \label{o5}
\ee
By performing a Poisson transformation one can also write the Kaluza-Klein
contributions as
\begin{equation}
F_d={{M_s^d}\over{(RM_s)^d}}\vartheta_3^d
\left(0,{i l (RM_s)^2 \over{2}} \right) \ . \label{o50}
\end{equation}
\subsection{The one loop amplitude as a sum of box diagrams}

We consider here in more detail the nonplanar diagram in the
representation given in (\ref{o4}).
The relevant exponentials of the Green functions (\ref{o1}) are 
explicitly given by
\be
\Psi= i e^{-\pi \tau_2 \nu^2} { \vartheta_1 (-i\nu \tau_2 , i \tau_2) \over
\eta^3}  \ , \
\Psi^{T}= e^{-\pi \tau_2 \nu^2} { \vartheta_2 (-i\nu \tau_2 , i \tau_2) \over
\eta^3} \ . 
\ee
In order to obtain a field theory interpretation of this string diagram,
it is convenient to first divide the region of integration over 
$\nu_i$ into three disjoint regions with a given ordering
\begin{equation}
R_1 \ : \nu_1<\nu_2<\nu_3<1 \ , \ R_2 \ : \nu_1<\nu_3<\nu_2<1 \ ,
\ R_3 \ : \nu_3<\nu_1<\nu_2<1 \ ,
\end{equation}
then we can write  
\begin{equation}
A=A^{(1)}(s,t)+A^{(2)}(u,t)+A^{(3)}(u,s) \ , 
\end{equation}
where\footnote{A similar parametrization for the four-point amplitude on
the torus in the Type II and heterotic strings was considered in
\cite{dhp}. There, the functions ${\cal R}^{(i)}$ are identical.} 
\be
A^{(i)}(s,t)\!=\!{8 g_{YM}^4 \over M_s^4}
\int_{0}^{\infty}d\tau_2 \ \tau_2 \int_{\eta_i>0} 
d^{4}\eta \delta(1-\sum_i\eta_i)
e^{{2\pi \tau_2 \over M_s^2}(s\eta_1\eta_3+t\eta_2\eta_4)}
\vartheta_3^6 (0,{{2i\tau_2}\over{(RM_s)^2}})
{ {\cal R}^{(i)}} \ . \label{o2}
\ee
The variables $\eta_i$ in the region $R_1$ are given by
\begin{equation}
\eta_1=\nu_1 \ ,\ \eta_i=\nu_i-\nu_{i-1} \ , \ i=2,3,4 
\end{equation}
and similar expressions in the other regions.
The factors ${\cal R}^{(i)}$ are given by
\begin{eqnarray}
{{\cal R}^{(1)}}
&=&
\left({{ f^{T}(\eta_3+\eta_2)
 f^{T}(\eta_3+\eta_4)}
\over{ f(\eta_2) f(\eta_4)}}\right)^{s/M_s^2}
\left({{ f^{T}(\eta_2+\eta_3) f^{T}(\eta_3+\eta_4)}
\over
{ f^{T}(\eta_1) f^{T}(\eta_3)}}\right)^{t/M_s^2} \ , \nonumber \\
{{\cal R}^{(2)}} &=&
\left({{f(\eta_3+\eta_2)
f(\eta_3+\eta_4)}
\over{f^T(\eta_2) f^T(\eta_4)}}\right)^{s/M_s^2}
\left({{ f(\eta_2+\eta_3) f(\eta_3+\eta_4)}
\over
{ f^{T}(\eta_1) f^{T}(\eta_3)}}\right)^{t / M_s^2} \ , \nonumber \\
{{\cal R}^{(3)}} &=&
\left({{f^T(\eta_3+\eta_2)
 f^T(\eta_3+\eta_4)}
\over{ f^T(\eta_2) f^T(\eta_4)}}\right)^{s/M_s^2}
\left({{ f^T(\eta_2+\eta_3) f^T(\eta_3+\eta_4)}
\over
{ f(\eta_1) f(\eta_3)}}\right)^{t/ M_s^2} \ , \label{o20}
\end{eqnarray}
where we introduced the convenient definitions
\be
f(\eta_i)=  e^{-\pi \tau_2 (\eta_i-1/6)} 
{ \vartheta_1 (-i \eta_i \tau_2,i \tau_2) \over \eta} \ , \
f^T(\eta_i)=  e^{-\pi \tau_2 (\eta_i-1/6)} 
{ \vartheta_2 (-i \eta_i \tau_2 , i \tau_2) \over \eta}
\ , \label{b1}
\ee
such that $f$ ($f^T$) can be expanded  in powers of
$exp(-2\pi \eta_i \tau_2)$ (B.10,B.11). 
By using the explicit definitions given in
Appendix A, it can be checked that $f(1-\eta_i)=f(\eta_i)$ and 
$f^T(1-\eta_i)=f^T(\eta_i)$, which was used in deriving (\ref{o20}). 
The field theory interpretation 
of $A$ is clarified by the formal expansion of
the factor ${\cal R}^{(i)}$ as a power series in 
$e^{-2\pi\tau_2\eta_j}$
\begin{equation}
{\cal R}^{(i)}=\sum_{n_1,\dots n_4\geq 0}
p^{(i)}_{n_1,\dots,n_4}\left({s \over M_s^2},{t \over M_s^2}\right)
e^{-2\pi\tau_2(n_1\eta_1+n_2\eta_2+n_3\eta_3+n_4\eta_4)} \ , \label{o3}
\end{equation}
where $p^{(i)}_{n_1,\dots,n_4}$ are polynomials in $s/M_s^2$
and $t/M_s^2$, whose explicit expressions are not important in
the following. Our definition is such that 
$p^{(i)}_{0,\dots,0}(s/M_s^2,t/M_s^2)=1$.

By using the expansion (\ref{o3}), the amplitude (\ref{o2}) can be
interpreted as a sum of an infinit set of box diagrams $B$ 
\begin{equation}
A^{(i)} = {12 g_{YM}^4 \over \pi^4 }
\sum_{k_1,\dots k_6=-\infty}^{+\infty}
\sum_{n_1,\dots n_4\geq 0}
p^{(i)}_{n_1,\dots,n_4}({s \over M_s^2},{t \over M_s^2})
B(s,t,\{n_iM_s^2\!+\!(k_1^2\!+\!\dots\!+\!k_6^2)/R^2 \}) \ . \label{bo}
\end{equation}
Indeed, the Feynman representation of a box diagram in 4D with particles
of masses $m_i^2$ in the loop reads
\ba
&&B(s,t,\{m_i^2\})=\int d^4 k \
\prod_{i=1}^{4}{{1}\over{(k+p_i)^2+m_i^2}} = \nonumber \\
&&{2 \pi^4 \over 3M_s^4 }
\int d\tau_2 \ \tau_2 d^4 \eta \ \delta (1-\sum_i \eta_i)
e^{{2\pi\tau_2 \over M_s^2} (s\eta_1\eta_3+t\eta_2\eta_4)}
e^{-{2\pi\tau_2 \over M_s^2} \sum_i m_i^2\eta_i} \ . \label{o6}
\ea
Note that in (\ref{bo}), the particles circulating in the loop 
are open string oscillators and KK states.

For $D\leq 4$ the box diagram with massless 
particles in the loop, which is the leading contribution to the above 
amplitude, is IR divergent. 
Infrared divergences are as usual harmless and in order to obtain a finite
intermediate result it suffices to add a small mass
to the particles circulating in the loop. Since $\sum_i \eta_i=1$,
it can be seen from (\ref{o6}) that this is equivalent to the
replacement
\begin{equation}
{\cal R}^{(i)}\rightarrow {\cal
R}^{(i)}e^{-2\pi\tau_2 \mu^2 \over M_s^2} \ ,
\end{equation}
where $\mu$ is a small mass which, as usual, is replaced by
the resolution over the energy of final particles in a given physical
process\footnote{Another way of regularizing the IR divergence is to
add a Wilson line in the Chan-Paton sector.}.

In practice, the expansion (\ref{bo}) 
is not very useful. It does not
correspond to an expansion in powers of
$s/M_s^2$ and $t/M_s^2$ and merely gives an interpretation 
of the non planar amplitude as an infinite sum of box diagrams.
The low energy limit is naturally given by the box diagram with
massless particles circulating in the loop.
However, this box diagram is UV divergent for spacetime dimension $D\ge
8$, whereas string theory is finite in the UV. This shows that this 
expansion is not manifesly UV finite. In fact for  $D\ge
8$ the series is divergent term by
term. This is to be contrasted with the expansion of
the similar torus amplitude in heterotic and Type II strings, where the
modular invariance provides an explicit UV cutoff \cite{dhp}. 
Furthermore, for $D\le 8$ , even though the diagrams are finite 
in the UV, infinitely many terms contribute to 
a given order in $s/M_s^2$. In section 3.4 we show how it is
possible to get a systematic low energy expansion of the amplitude,
which is also manifestly UV finite. Before doing
that, we will need however another interpretation of the non planar
diagram. 
\subsection{The nonplanar  amplitude as exchange of closed string
modes}

In this subsection we shall discuss a representation which
is the string generalisation of
the proper time parametrisation used in equation (\ref{I2}).
In this representation we must express the amplitude (\ref{o5}) 
as a function of
$l=\tau_2^{-1}$, with $l$ the modulus of the cylinder. 
The factor $F_6$ due to the Kaluza-Klein modes  is given by (\ref{o50}),
while the functions $\tilde\Psi$, $\tilde\Psi^T$ defined in
(\ref{psi}) are given by
\be
\tilde\Psi = {1 \over l} {\vartheta_1(\nu,il) \over \eta^3} \ , \
\tilde\Psi^{T} =  {1 \over l} {\vartheta_4(\nu,il) \over \eta^3} \ . \label{c2}
\ee
In order to analyse this amplitude it is convenient to define
\begin{equation}
 \left({{\tilde\Psi_{13}^{T}\tilde\Psi_{24}^{T}}
\over{\tilde\Psi_{12}\tilde\Psi_{34}}}\right)^{s/M_s^2} \!\!
\left({{\tilde\Psi_{13}^{T}\tilde\Psi_{24}^{T}}\over
{\tilde\Psi_{14}^{T}\tilde\Psi_{23}^{T}}}\right)^{t / M_s^2}
\!\!\!\!\!\!= e^{\pi l s /2M_s^2 } \left[4\sin{\pi(\nu_2\!-\! \nu_1)}
\sin{\pi(1\!-\! \nu_3)}\right]^{-s/M_s^2}\tilde{\cal
R} \left( {s \over M_s^2}, {t \over M_s^2},\nu_i \right) \ ,
\end{equation}
where
\begin{equation}
\tilde{\cal R}\equiv\left({{\tilde f_{13}^{T}\tilde f_{24}^{T}}
\over{\tilde f_{12}\tilde f_{34}}}\right)^{s/M_s^2}
\left({{\tilde f_{13}^{T}\tilde f_{24}^{T}}\over
{\tilde f_{14}^{T}\tilde f_{23}^{T}}}\right)^{t / M_s^2}
\end{equation}
and
\be
\tilde f(\nu,l)= {e^{\pi l /6} \over 2 \sin{\pi \nu}}
{\vartheta_1 (\nu,il) \over \eta} \ , \
\tilde f^{T}(\nu,l) = e^{-\pi l /12} {\vartheta_4 (\nu,il) \over \eta} \ .
\ee
Similarly to $f$ and $f^T$ in (\ref{b1}), these functions can be
expanded in positive powers of $e^{-2\pi l}$. Therefore
it is possible to cast $\tilde{\cal R}$ in the form
\begin{equation}
\tilde{\cal R}=1+\sum_{n=1}^{\infty}
\tilde p_n(s/M_s^2,t/M_s^2,\nu)e^{-2\pi n l} \ , 
\end{equation}
where $\tilde p_n$ is a polynomial in $s/M_s^2$, $t/M_s^2$ and
$e^{2i\pi\nu}$, whose exact expression is not important here.
Assembling the different terms and by finally using (\ref{coup}), the
amplitude can be cast in the suggestive form
\begin{equation}
A={1 \over M_P^2 M_s^2}
\sum_{n=0}^{\infty} c_n \int_{0}^{\infty}dl
\sum_{n_1,\dots n_6}
e^{-\pi l/2[-s/M_s^2+(n_1^2+\dots+n_6^2)(RM_s)^2+4n]} \ , \label{c3}
\end{equation}
where we defined 
\begin{equation}
c_n= \int d\nu_1 d\nu_2 d\nu_3 \ {\tilde p}_n(s/M_s^2,t/M_s^2,\nu)
\left[ 4\sin{\pi(\nu_2-\nu_1)}
\sin{\pi(1-\nu_3)} \right]^{-s/M_s^2} \ ,
\end{equation}
for $n \geq 0$. The field theory result (\ref{I1}) is obtained by truncating in
(\ref{c3}) the massive string oscillators and
taking the low energy limit in $c_0$, which gives $1/2$. We therefore
keep only the winding modes, T-dual to the KK states appearing in (\ref{I1}).
 
The integral in (\ref{c3}) is simply the proper time 
representation of a Feynman propagator with the mass 
\begin{equation}
m^2= M_s^2 \ [(n_1^2+\dots+n_d^2)(RM_s)^2+4n] \ . \label{mass}
\end{equation}
This fact reflects a familiar result: the one loop open string amplitude  
can be seen as a tree diagram in the closed string channel
\cite{dua}
where the masses of the closed string particles are given by
(\ref{mass}), the integer $2n$ being the closed string oscillator level and 
$n_1,\dots, n_6$ the winding numbers.
Note however that the expansions we performed  and therefore a
truncation for some value on $n$ are valid for large $l$. 
Similarly to the case of the representation of the non planar
amplitude as a sum of box diagrams, this representation
is not manifestly UV convergent. In fact the sum over the
winding modes behaves as $l^{-d/2}$ for small $l$, so that the integral
diverges for $d \geq 2$ and in particular in the present case $d=6$ we
obtain a quartic divergence.
\subsection{Type I ultraviolet regularisation of
10D field theory} 

In the two preceding sections we have given two representations of
the non planar amplitude. Both of them were not manifestly UV convergent
and did not allow a systematic (in $s/M_s^2$ and $t/M_s^2$) low 
energy expansion. Here we combine both of them in a new representation
which is free of these two drawbacks.

The two dual expressions (\ref{o4}) and (\ref{o5}) are typically of
the form
\begin{equation}
I=\int_{0}^{\infty}dx \ h(x) [g(x)]^\epsilon \ ,
\end{equation}
and we are interested in the small $\epsilon$ expansion of
the amplitude. In the easiest case where $g$ is strictly positive and 
bounded from above, the expansion is given by expanding the integrand,
that is
\begin{equation}
I=\int_0^\infty dx \ h+\epsilon\int_0^\infty dx \ h(x)\ln{(g(x))}+\dots \ .
\end{equation}
A less trivial case is when $g$ vanishes
somewhere between $0$ and $\infty$ and possibly at 
the boundaries, in which case one cannot perform an expansion of the
above form.
Suppose however that $g$ can be put in the form
$g=g_1(x)g_2(x)$ where $g_2$ is strictly positive
and bounded. Then one can expand $I$ as
\begin{equation}
I=\int_0^\infty dx \ h[g_1]^\epsilon+\epsilon \int_0^{\infty} dx \ h
[g_1]^\epsilon\ln(g_2)+\dots \ ,
\end{equation}
which can be useful when $g_1$ is much simpler than $g$.
 
\noindent Now we come back to the string amplitude, which
in the open string representation has the form
\begin{equation}
A=\int_{0}^{\infty}{{d\tau_2}\over{\tau_2^2}}\int d^4\eta
\delta(1-\sum_i \eta_i) e^{{2\pi\tau_2 \over M_s^2} (s\eta_1\eta_3 + t\eta_2\eta_4) }
I^{s/M_s^2} J^{t/M_s^2} \ , 
\end{equation}
where both $I$ and $J$ are bounded and nonvanishing at $\tau_2=\infty$.
However at $\tau_2=0$ they vanish and furthermore it is not
possible to factorise a finite number of vanishing terms.
The closed string representation of the amplitude 
has the same problem : it is possible to isolate the dangerous
piece
at $l=\infty$ but not at $l=0$. 
In fact the closed string representation
can be put in the form
\begin{equation}
A=\int_{0}^{\infty} dl \ e^{\pi l s / 2M_s^2}\int d^3\nu \ 
[4\sin\pi(\nu_2\!-\! \nu_1)\sin\pi(1\!-\! \nu_3)]^{-s/M_s^2}\tilde
I^{s/M_s^2}\tilde J^{t/M_s^2} \ ,
\end{equation}
where $\tilde I$ and $\tilde J$ are bounded and nonzero at
$\infty$ but not at $l=0$. Note that the problematic 
region of each representation corresponds to the nice region of the
other represention. This suggests a solution which consists
in using a mixed representation of the amplitude :
choose a finite nonvanishing $l_0$ and write the 10D amplitude as
\begin{eqnarray}
A &=& {g_s^2 \over M_s^{10}} \int_{1/l_0}^{\infty}{{d\tau_2}\over{\tau_2^2}}\ 
\int_{0}^{1}d\nu_2\int_{0}^{\nu_2}d\nu_1
\int_{0}^{1}d\nu_3\ 
\left({{\Psi^T_{13}\Psi^T_{24}}
\over{\Psi_{12}\Psi_{34}}}\right)^{s/M_s^2}
\left({{\Psi^T_{13}\Psi^T_{24}}
\over{\Psi^T_{14}\Psi^T_{23}}}\right)^{t/M_s^2} \nonumber \\
&+& {g_s^2 \over M_s^{10}} \int_{l_0}^{\infty}dl\ 
\int_{0}^{1}d\nu_2\int_{0}^{\nu_2}d\nu_1
\int_{0}^{1}d\nu_3\ 
\left({{\tilde\Psi^T_{13}\tilde\Psi^T_{24}}
\over{\tilde\Psi_{12}\tilde\Psi_{34}}}\right)^{s/M_s^2}
\left({{\tilde\Psi^T_{13}\tilde\Psi^T_{24}}
\over{\tilde\Psi^T_{14}\tilde\Psi^T_{23}}}\right)^{t/M_s^2} \nonumber \\
&\equiv& A_1(l_0)+A_2(l_0) \ .
\end{eqnarray}
Now in each integral we can use the factorisation 
described above.
Note that even if the full amplitude is independent on $l_0$,
each part of it clearly does depend. In fact
$l_0$ plays the r\^ole of an UV cutoff $\Lambda_c$
for the closed string exchange
and $l_0^{-1}$ plays the r\^ole of an UV cutoff $\Lambda_o$
in the one loop box diagrams. The two cutoffs are given by
\begin{equation}
\Lambda_c={{M_s}\over{\sqrt{l_0}}} \quad , \ \Lambda_o=M_s\sqrt{l_0}  
\end{equation} 
and are clearly inversely proportional to each other:
\begin{equation}
\Lambda_o\Lambda_c=M_s^2 \ .
\end{equation}
This mixed representation is thus manifestly UV
convergent. The low energy expansion is also
manifestly finite term by term.
 
Let's elaborate more on this 10D example and obtain the finite result
for low energy limit of the amplitude.
Consider first the $A_1(l_0)$ part
\begin{eqnarray}
A_1(l_0)={g_s^2 \over M_s^{10}} \int_{1/l_0}^{\infty}{{d\tau_2}\over{\tau_2^2}}
\int d^4\eta \delta(1-\sum_i \eta_i)
e^{{2\pi\tau_2 \over M_s^2} (s\eta_1\eta_3+t\eta_2\eta_4)}\nonumber\\
 \Big(1+{s \over M_s^2} \ln I + {t \over M_s^2} \ln J +\dots\Big) \ .
\end{eqnarray}
Let us first neglect the terms multiplying $s/M_s^2$
and $t/M_s^2$, as well as higher order terms
and integrate over $\tau_2$ to obtain
\begin{equation}
A_1(l_0)={g_s^2 \over M_s^{10}} l_0 \int d^4\eta \delta(1-\sum_i \eta_i)E_2
(-{2\pi \over l_0 M_s^2} (s\eta_1\eta_3+t\eta_2\eta_4))+\dots
\end{equation}
where
\begin{equation}
E_m(z)=\int_{1}^{\infty}\ {{dx}\over{x^m}}{{e^{-zx}}} \ . \label{Em}
\end{equation}
For small $z$ we have
\begin{equation}
E_2(z)=1+z\ln z-(1-\gamma)z+\dots
\end{equation}
so to the next to leading order in $s/M_s^2$ we have
\begin{equation}
A_1^{(1)}(l_0)={{l_0} g_s^2 \over 6M_s^{10}}-{2\pi g_s^2 \over M_s^{12}}
\int d^4\eta \delta(1\!-\!\sum_i \eta_i)(s\eta_1\eta_3+t\eta_2\eta_4))
\ln {-(s\eta_1\eta_3+t\eta_2\eta_4) \over M_s^2} +\dots \label{t1}
\end{equation}
if we neglect higher order terms. Note that the leading term 
depends on $l_0$ and becomes infinite in the $l_0\rightarrow \infty$
limit, which signals that the 10D Yang-Mills 
box digram is UV divergent. The $l_0$ dependence
must of course cancel in the full amplitude. In order to check it
explicitly let us consider the second part of the amplitude
\be
A_2(l_0)\!\!=\!\!{g_s^2 \over M_s^{10}} \int_{l_0}^{\infty} dl e^{\pi l
s / 2 M_s^2} \! \int d^3\nu
[4\sin\pi(\nu_2\!-\!\nu_1)\sin\pi(1\!-\!\nu_3)]^{-s/M_s^2}
 \{1\!+\!\!{s \over M_s^2} \ln\tilde I\!+\!\!{t \over M_s^2} \ln\tilde
J+\dots\} \ . \label{t2}
\ee
The $s/M_s^2$ and $t/M_s^2$ terms in the brackets, representing
oscillator contributions, give a vanishing contribution to the
first order in $s/M_s^2$.
In fact
\begin{equation}
\int d^3\nu \ln {\tilde I}=\int d^3\nu \ln {\tilde J}=0 \ ,
\end{equation}
as can be easily verified upon expanding the logarithms
in powers of $e^{2\pi i\nu}$. So to the first order 
in $s/M_s^2$ or, more generally, if we neglect string oscillator
exchanges in $A_2$, we can replace the terms in the curly brackets by $1$.
The integral over $\nu$ can then be performed and gives
\begin{equation}
\int d^3\nu
[\sin\pi(\nu_2-\nu_1)\sin\pi(1-\nu_3)]^{-s/M_s^2}
={{1}\over{2\pi}}
\left({{\Gamma\left(-s/2M_s^2+1/2 \right)
\sqrt{\pi}}
\over{\Gamma\left(-s/2M_s^2+1\right)}}\right)^2 \ . \label{t3}
\end{equation}
This result,  due to the massless graviton tree-level exchange in 10D,
presents a perfect square structure which allows the identification of
the tree-level (disk) form factor $g$ between two (on-shell) gauge bosons and
one (off-shell) massless graviton of momentum squared $s$ to be
\be
g (s) = {1 \over \sqrt{\pi}} 2^{-{s \over M_s^2}} { \Gamma \left(-s/2M_s^2+1/2 \right)
\over \Gamma \left(-s/2M_s^2+1\right)} \ . \label{t4}
\ee
The presence of poles (and also zeroes) in this form factor is
interpreted as due to a tree-level mixing between the massless graviton
and open string
singlets, present at odd mass levels in the $SO(32)$ Type I
superstring. The result (\ref{t4}) will be rederived and generalized for
off-shell gauge bosons and compactified 4D theory in Section 4. 

To the first order in $s/M_s^2$, (\ref{t3}) becomes
$1 / 2+s/M_s^2 \ 2\ln2+\dots$.
When combined with $4^{-s/M_s^2}$, the first order terms in $A_2$ in
$s/M_s^2$ cancel and we are left with
\be
A_2(l_0)={-g_s^2 \over \pi s M_s^8 } (e^{\pi s l_0 / 2M_s^2}+
O((s^2/M_s^4)) ={-1 \over \pi s M_P^8}-{l_0 g_s^2 \over 2 M_s^{10}} +
\cdots \ ,
\ee
where in identifying the graviton pole in the last equality we used the
first relation in (\ref{coup}) for 10D (d=0). 
To order $(sg_s^2/M_s^{12}) \ln (s/M_s^2) $, the amplitude is then given by
\begin{equation}
A=A_1^{(1)}(l_0)(s,t)+A_1^{(1)}(l_0)(u,t)+A_1^{(1)}(l_0)(u,s)
+A_2(l_0) \ .
\end{equation}
Notice first of all that up to this order the terms dependent on $l_0$ cancel
in the sum $A_1(l_0)+A_2(l_0)$, as it should. 
The first term in the r.h.s. of $A_2$ represents 
the exchange of the graviton multiplet between the two gauge
bosons. One may be tempted to interpret $A_2$
as the gravitational contribution to the amplitude.
This is unambigous as long as we are considering the leading
contributions in $A_1$ and $A_2$. However,
if we consider higher order terms the distinction
between $A_1$ and $A_2$ looses its relevance.
In other words the gauge and gravitational contributions 
are mixed and only their sum in meaningful. 
Notice also the absence of terms of
order $0$ in $s/M_s^2$ in $A$. This is a direct verification of the
absence of one loop $(trF^2)^2$ terms in the effective action of
the type I superstring, a result which was important 
in checking the type I/heterotic duality \cite{ts}.

Let us consider the effect of including higher order terms.
The dominant contribution of $\Delta A_1$ is of order
$s g_s^2/M_s^{12}$. However this contribution is $l_0$ dependent
and thus must cancel with a similar contribution from $A_2$.
The first $l_0$ independent contribution to $\Delta A_1$
is of order $(g_s^2 s^2/M_s^{14}) \ln(-s/M_s^2)$, which can be easily
checked explicitly.

Notice that in this way, one sees clearly
how type I strings regularise the UV divergent field
theory. In field theory language, this is equivalent to
introducing an {\it arbitrary} UV cutoff in the divergent (box) diagram
and using a related cuttoff in the graviton exchange one. The 
product of the two is $M_s^2$ and the sum of the two diagrams is cutoff-independent.
One may interpret the result as a regularisation
of the Yang-Mills theory by gravity. Notice that this works in a
subtle way because the gravity diagram here is UV finite.
The regularisation is possible because the cutoff used on the
gravitational side is inversely proportional to the one used for
the Yang-Mills diagram.

\subsection{Type I ultraviolet regularisation of winding
modes in 4D}

In 4D the box diagrams are UV convergent but the closed string exchange 
is UV divergent. 
Introducing the parameter $l_0$ and using a mixed representation we get a
manifestly UV finite form of the amplitude.
The IR divergence of the box diagram is eliminated
by adding a small mass $\mu$ to the open string modes or, in a SUSY
manner, by adding a Wilson line in the open sector.

Tha amplitudes can be expanded in $RM_s$ as well as in
$s/M_s^2$ and $t/M_s^2$.
Since we are interested in the small $RM_s$ limit we shall keep
the leading contribution in $RM_s$ and the next to leading
contribution in $s/M_s^2$.
The leading contribution in $(RM_s)$ in $A_1$ is obtained
by neglecting all the Kaluza-Klein modes whose relative
contributions are of order $(RM_s)^4$. In $A_2$
the sum over the winding modes is replaced by the zero mode
and an integral over non-zero modes, that is
\begin{equation}
\sum_{n_1,\dots n_6}
e^{-\pi l[-s/2M_s^2+(n_1^2+\dots+n_6^2)(RM_s)^2/2+2n]}
=e^{\pi l s/2M_s^2}(1+{{8}\over{(RM_s)^6 l^3}})+\dots.
\end{equation}
Note that strictly speaking one should neglect the zero mode
contribution, however this is the only term that diverges in the
$s/M_s^2\rightarrow 0$ limit. The $A_2(l_0)$ contribution becomes
\begin{eqnarray}
A_2(l_0)&\!\!\!=\!\!\!&{1 \over M_s^2 M_P^2}
\int_{l_0}^{\infty} dl e^{\pi l s/2M_s^2} (1\!+\!{{8}\over{(RM_s)^6 l^3}})
\int d^3\nu [4\sin\pi(\nu_2\!-\! \nu_1)\sin\pi(1\!-\! \nu_3)]^{-s/M_s^2}\nonumber\\
&& \{1+{s \over M_s^2} \ln\tilde
I+{t \over M_s^2} \ln\tilde J+\dots\} \ . \label{f1}
\end{eqnarray}
As in the previous Section, eq. (\ref{t2}), if we neglect string
oscillator contributions coming from ${\tilde I},{\tilde J}$, the $\nu$ 
integral gives an effective form factor (\ref{t4}), which is seen now to
be the same for {\it all} winding states, result which will be rederived and
shown to be true even for off-shell gauge bosons in Section 4. 
 
As in the 10D case the dominant contribution in $A_2$ comes from
\begin{equation}
A_2^{(0)}(l_0)={1 \over 2 M_s^2 M_P^2} \int_{l_0}^
{\infty} dl \ e^{\pi l s/2M_s^2} \left(1+{{8}\over{(RM_s)^6 l^3}} \right) \ ,
\end{equation}
where the factor $1/2$ comes from the integration over $\nu$.
This contribution can be expressed with the aid of the $E_3$ function 
(\ref{Em}) as
\begin{equation}
A_2^{(0)}(l_0)=- {1 \over \pi s M_P^2} e^{\pi l_0 s/2M_s^2 }
+{4 \over l_0^2 M_P^2 R^6 M_s^8} E_3(-{\pi l_0 s \over 2M_s^2}) \ . \label{aaa}
\end{equation}
The small $s/M_s^2$ limit (\ref{aaa}) is obtained 
with the aid of the developpement of $E_3$
\begin{equation}
E_3(z)={{1}\over{2}}
\left(1-2z-z^2\ln z+({{3}\over{2}}-\gamma)z^2+\dots\right),
\end{equation}
where $\gamma$ is the Euler constant. Therefore we find, by using again (\ref{coup})
\be
A_2^{(0)}(l_0)\!=\! -{1 \over \pi s M_P^2} + {2 g_{YM}^4 \over M_s^4}
\left({{1}\over{l_0^2}}+{\pi s \over l_0 M_s^2}
-{\pi^2 s^2 \over 4 M_s^4} \ln(-{\pi l_0 s \over 2M_s^2})
+({3 \over 2}\!-\! \gamma) {\pi^2 s^2 \over 4 M_s^4} +\dots\right) \ . \label{f2}
\ee
It is transparent in (\ref{f2}), because of the factor $g_{YM}^4$, that 
the leading corrections to the graviton exchange diagram are actually 
mostly related to the one-loop box diagram and not to massive tree-level
exchanges. However, as already emphasized, the real physical quantity is
the sum of $A_1$ and $A_2$.
  

We now turn to the $A_1(l_0)$ contribution,
where we rely heavily on technical results derived in 
Appendix B.
The dominant contribution is obtained from
\begin{equation}
 A^{(i,0)}_1 (s,t)={8 g_{YM}^4 \over M_s^4}
\int_{1/l_0}^{\infty}d\tau_2 \tau_2 \int_{\eta_i>0} 
d^{4}\eta \delta(1-\sum_i\eta_i)
e^{{2\pi \tau_2 \over M_s^2}(s\eta_1\eta_3+t\eta_2\eta_4)}
e^{-{2\pi\tau_2 \mu^2 \over M_s^2}} \ .
\end{equation}
Since the box diagram is UV convergent, we can safely
write the above integral as
\begin{equation}
A^{(i,0)}_1 (s,t)=\int_0^\infty...-\int_{0}^{1/l_0}... \ .
\end{equation}
The first term is obtained from the box diagram
calculated in the Appendix B 
\begin{equation}
\int_0^\infty...= {2 g_{YM}^4 \over \pi^2 }
{{1}\over{st}}\ln{{-s}\over{4\mu^2}}\ln{{-t}\over{4\mu^2}} \ 
\end{equation}
and the second is given by the following expansion 
in $s/M_s^2$ and $t/M_s^2$
\be
\int_{0}^{1/l_0}...\!=\!{8 g_{YM}^4 \over M_s^4}
\int_{0}^{1/l_0}d\tau_2 \tau_2\int_{\eta_i>0} 
d^{4}\eta \delta(1\!-\!\sum_i\eta_i)
[1\!+\!{2\pi \tau_2 \over M_s^2} (s\eta_1\eta_3\!+\!t\eta_2\eta_4)
+ \dots ] \ .
\ee
These integrals can be easily evaluated and yield
\begin{equation}
\int_{0}^{1/l_0}...={8 g_{YM}^4 \over M_s^4}
[{{1}\over{12l_0^2}}+{\pi (s+t) \over 180 l_0^3M_s^2}] + \cdots \ .
\end{equation}
Note that the second term when considering
the sum $A^{(1)}+A^{(2)}+A^{(3)}$ gives a vanishing contribution
due to $s+t+u=0$. The next contribution to $A_1$ comes from box diagrams 
with a massive string mode in one propagator, the other three
propagators containing light particles (of mass $\mu$).  
As before, since the diagrams are UV convergent,
in order to get the $l_0$ independent terms it suffices to calculate 
the corresponding box diagram. These diagrams are calculated 
in  Appendix B. The $l_0$ independent terms in the sum of the three terms
are 
\be
\Delta A_1 = -{g_{YM}^4 \over 3M_s^4}
\left(\ln {{s}\over{t}}\ln {{st}\over{\mu^4}}
+\ln {{s}\over{u}}\ln {{su}\over{\mu^4}}\right) \ .
\ee
Notice that the terms in $(\ln \mu^2)^2$ have cancelled 
in the sum of the three terms in (\ref{cor}).
Collecting terms of lowest order in $s/M_s^2$
the contribution of $A_1$ reads
\ba
&&\!\!\!\! A^{(1,0)}(s,t)+A^{(1,0)}(u,t)+A^{(1,0)}(u,s)+
\Delta A_1 =  \nonumber \\
&&\!\!\!\! {2 g_{YM}^4 \over \pi^2} \ [ \ {1 \over st} \ln {-s \over
4 \mu^2} \ln {-t \over 4 \mu^2}\!+\! {\rm perms.}] - 
{g_{YM}^4 \over 3 M_s^4} 
\ [ \ln {s \over t}  \ln {s t \over \mu^4} \!+\! \ln {s \over u} \ln {s u
\over \mu^4} \!+\! {6 \over l_0^2}] \!+\! \cdots \ , \label{f4}
\ea
the result announced in (\ref{I7}).
The $l_0$ dependent terms to this order are
$(2 g_{YM}^4 / l_0^2 M_s^4)$ which exactly cancels the first $l_0$
dependent term in $A_2$ as it should.

The next corrections in $s/M_s^2$ to $A_1$ 
come from terms of the type $(\ln f(\eta_2))^2$
and from terms of the type $\ln f(\eta_2+\eta_3)$, which
represent a sum of box diagrams with two massive modes circulating in the loop.
By using the result (B.8) in Appendix B, 
it can be shown that the
corresponding corrections to (\ref{f4}) are of the order
$g_{YM}^4 s M_s^{-6}\ln M_s^2/\mu^2$.
\section{Couplings of brane states to bulk states}

Another interesting computation, closely related to the tree-level
exchange of virtual closed string states
is the tree-level (disk) coupling between two open (brane) states and
one closed (bulk) winding excitation of mass $w^2 \equiv {\bf n}^2 R^2 M_s^4$,
where ${\bf n}^2 = n_1^2+ \cdots n_6^2$. We show here, in agreement with
the results obtained in Sections 3.4 and 3.5, that {\it all} winding modes couple the same
way to the gauge bosons with a form factor written in (\ref{t4}).
Recently this issue was investigated in an effective theory context
\cite{bkny} and an exponential supression in the winding (KK after
T-duality) modes was found, interpreted there as the brane thickness.
In a field theory context, the result depends not only on the
fundamental mass scale but also on other (dimensionless) parameters. The
full result has a nonperturbative origin from string theory viewpoint. In
the perturbative string framework we discuss here, the result depends
only on the string scale $M_s$. The form factor that we found
in (\ref{t4}) depends actually on the energy squared of the graviton and
not on its mass, a difference which is important for off-shell calculations.
Moreover, its presence is actually, as discussed in detail in previous
Sections, not directly related to the regularization of winding (KK)
virtual summations.
 
\subsection{Two gauge bosons - one winding graviton amplitude} 

We consider for illustration the case of the open bosonic string. A
similar computation in the superstring case was performed in \cite{HK}
and we compare here their result with ours in order to understand the
role played by supersymmetry in these computations. We
compute the correlation function of two gauge boson vertex operators $V_i^{\mu_i}$
with one one bulk graviton vertex operator $V_3^{\mu_3
\mu_4}$ on the disk represented here as the upper half complex plane
$z$, $Im z >0$. We use the doubling trick to represent the
antiholomorphic piece of the graviton vertex operator as an holomorphic
operator at the point $z'={\bar z}$. Then the vertex operators are of the form 
\ba
&&V_i^{\mu_i}=g_s^{1/2} \lambda^{a_i} :\partial X^{\mu_i} (y_i) e^{2ip_i.X(y_i)}:
\ , \nonumber \\  
&&\!\!\!V_3^{\mu_3 \mu_4}\!=\!V_3^{\mu_3}(z) V_3^{\mu_4}({\bar z})\!= \! 
g_s :\partial X^{\mu_3}(z)e^{{ip_3 }X(z)\!+\!iwY(z)}: 
:\partial X^{\mu_4}({\bar z}) 
e^{{ip_3 }X({\bar z})\!-\! iwY(\bar z)}:  
\ , \label{co1}
\ea
where $X$ are spacetime coordinates, $Y$ compact coordinates and  
$\lambda^{a_i}$ Chan-Paton factors gauge bosons.
The gauge vector vertex operators are inserted on boundary points $y_1,y_2$
and the graviton vertex operator on a bulk point $z$. The Green
functions on the disk we need in the computation are
\be
<X^{\mu}(z_1) X^{\nu}(z_2)>\!=\! -{1 \over 2 M_s^2} \eta^{\mu \nu} \ln (z_1\!-\!z_2) \ , \
<X^{\mu}(z_1) X^{\nu}({\bar z_2})>\!=\! -{1 \over 2 M_s^2} \eta^{\mu \nu} 
\ln (z_1\!-\! {\bar z_2}) \ , \label{co2}
\ee
where $\eta^{\mu \nu}$ is the Minkowski metric. The disk has three
conformal Killing vectors which allow to fix three parameters in the positions of the
vertex operators. We choose to fix the position of the graviton and the
position of the second gauge boson. Introducing polarization vectors
$\e_i$ for gauge bosons and $\e_3$ for the graviton, the amplitude to
consider is then
\be
A_{ab} = g_s^{-1} N_3 tr (\lambda_a
\lambda_b) \int_{-\infty}^{\infty} dy_1
<c(y_2)c(z)c({\bar z})> < \epsilon_1 V_1(y_1) \epsilon_2
V_2 (y_2) \epsilon_3 V_3 (z) \epsilon_3 V_3 ({\bar z})>
\ , \label{co3}
\ee
where $<c(y_2)c(z)c({\bar z})>=|(y_2-z)(y_2-{\bar z})(z-{\bar z})|$ is the factor coming from
fixing the three positions on the disk, the factor $g_s^{-1}$ comes from
the topological factor of the disk and $N_3$ is a normalization constant
to be fixed by factorization later on. The conditions of transverse
polarizations are $p_1 \e_1=p_2 \e_2 =0$, $p_3 \e_3=0$
and the graviton polarization is also constrained by eliminating the
scalar component $\e_3^2=0$. Is to be understood that in the final
result, $\epsilon_3^{\mu} \epsilon_3^{\nu}$ is to be replaced by the
symmetric polarization tensor of the graviton,  $\epsilon_3^{\mu \nu}$.

The mass-shell conditions are $p_1^2=p_2^2=0$, $p_3^2=-w^2$ and
the kinematics of the process is described by
\ba
s &=&-(p_1+p_2)^2= w^2 \ , \ p_1 p_2= -{1 \over 2} w^2 \ , \nonumber \\  
p_1 p_3 &=& {1 \over 2} w^2 \ , \  p_2 p_3= {1 \over 2} w^2 \ . \label{co4}
\ea
The gauge choice we make in the following is $y_1=y$, $y_2=0$ and
$z=i$. Then, by a straightforward computation of (\ref{co3}) and by using the
formula
\be  
B(a,b)= \int_{-\infty}^{\infty} y^{2a-1} (1+y^2)^{-(a+b)} =
{\Gamma(a) \Gamma(b) \over \Gamma(a+b)} \ ,
\label{co5}
\ee      
where $\Gamma(a)$ is the Euler function, we find the final result for the amplitude   
\ba
A_{ab}&\!=\!& {{4}\over{\sqrt \pi M_P}}
2^{-{w^2 \over M_s^2}} \{ \bigl[ (2\e_1 \e_2) (p_1 \e_3)^2\!-\!
(\e_1 \e_3) (\e_2 \e_3) \!+\! 8 (\e_1 p_2) (\e_2 p_1) (\e_3 p_1)^2
\bigr] B(\!-\!{s^2 \over 2 M_s^2}\!+\!{3 \over 2}, {1 \over 2}) \nonumber \\
&&+ \bigl[ 2 \e_1 \e_2 (p_2 \e_3)^2 + (\e_1 \e_3) (\e_2 \e_3) + 4
(p_1\e_2)(p_2\e_3)(\e_1\e_3)+ \nonumber \\ 
&&4 (\e_1p_2)(\e_2\e_3)(p_1\e_3)+ 8 (\e_1p_2)(\e_2p_1)(\e_3p_2)^2 \bigr] 
B(-{s \over 2 M_s^2}+{1 \over 2}, {3 \over 2}) \} \delta_{ab} \ , \label{co6} 
\ea
where $N_3$ was determined by factorization of the one-loop four-point
amplitude of Section 3.5.
Notice that the amplitude has a sequence of poles for $s=
w^2=(2n-1)M_s^2$, with $n=1 \cdots \infty$ a positive integer, and zeroes
for $s=2(n+1)M_s^2$. The poles can be interpreted as due to massive 
open string states at odd levels coupled to the gauge fields and to the massive
graviton through a tree-level diagram, as in Figure 6. In order
for this to be possible, these states must be gauge singlets. Indeed, in the
toroidal compactification we are considering, the gauge group is
orthogonal ($SO(2^{13})$ for the bosonic string) and the spectrum
contains adjoint (antisymmetric) representations at even mass levels and
symmetric representations at odd mass levels. The symmetric
representations however are reducible and contain the singlets which
produce the poles. Note that, even if the spectrum at
odd mass levels starts with a tachyonic state, this does not couple and
therefore produce no pole in the amplitude. The particular case $w^2=0$
of the amplitude is in agreement with the field-theoretical
computation of the three-point amplitude computed from the interaction Lagrangian
\be
{\cal L} = {1 \over M_P} h_{\mu \nu} T^{\mu \nu}+ \cdots = 
{1 \over M_P} h_{\mu \nu} \ tr \ \bigl( F^{\mu \rho} F_{\rho}^{\nu} - {1 \over 4} 
\eta^{\mu \nu} F_{\rho \sigma} F^{\rho \sigma} + \cdots \bigr) 
\ , \label{co7}
\ee 
where $T^{\mu \nu}$ is the energy-momentum tensor of the gauge-fields
and $h_{\mu \nu}$ represents the graviton. More precisely, an explicit
3-point computation from (\ref{co7}) reproduces exactly all terms 
in (\ref{co6}) except the terms quartic in momenta. Up to these terms
actually the result is exactly the same as in the superstring
case\footnote{This can be shown by using the identity $2^{2z-1} 
\Gamma(z) \Gamma(z+1/2) = \sqrt{\pi} \Gamma (2z)$.}
\cite{HK} and therefore the conclusions we present below are largely
independent on supersymmetry\footnote{This allowed us to determine the
normalisation constant $N_3$ in (\ref{co3}) by factorizing the one-loop 
amplitude of Section 3.5.}. 
In particular, we find here again the
selection rule which make the amplitude vanish for  $s=2(n+1)M_s^2$.
The quartic terms, absent in the
superstring case, are to be interpreted as arising from
the higher-derivative term in the Lagrangian $F^{\rho \mu}F^{\sigma \nu}
R_{\rho \mu \sigma \nu}$, with $R_{\rho \mu \sigma \nu}$ the
gravitational Riemann tensor. 

\begin{figure}
\vspace{4 cm}
\special{hscale=60 vscale=60 voffset=0 hoffset=120
psfile=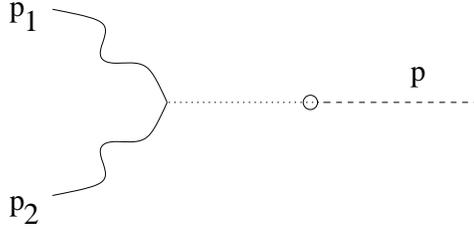}
\caption{The two gauge bosons -- one graviton vertex, 
where the intermediate states are open string singlets.}
\end{figure}

The computation presented is on-shell $s=w^2$. We are now
interested in the large s=$w^2$ behaviour of the above
amplitude (for values which avoid the poles and zeros we just
discussed), which on-shell is equivalent of considering couplings to very
massive winding gravitons. By using the asymptotic expansion (valid for $s >> M_s^2$)
\be
B(-{s \over 2 M_s^2}+{3 \over 2}, {1 \over 2}) \simeq  
\sqrt{2\pi M_s^2 \over s} \tan{\pi s \over 2M_s^2} \ ,  
\ee
we find the (on-shell) effective coupling of gauge fields to massive winding (or
Kaluza-Klein in the T-dual picture where the gauge field is stuck on a
D3 brane orthogonal to the compact space)
\be
g_{\bf n} = {1 \over \sqrt{\pi}} 2^{-w^2 \over M_s^2}{\Gamma
(-w^2/2M_s^2+1/2) \over \Gamma (-w^2/2M_s^2+1)} \sim  2  
\sqrt{2M_s^2 \over \pi w^2} 
\tan{\pi w^2 \over M_s^2} e^{-{w^2 \over M_s^2} \ln 2} \ , \label{co8}            
\ee
where in the last formula we took the heavy mass limit $w^2 >> M_s^2$. 
So, modulo the power in front of it, we find an exponential supression
of states heavier than a cutoff mass $\Lambda^2 =M_s^2/\ln 2$. However,
we emphasize again that this interpretation is valid for masses not very
close to poles and zeros of the full expression (\ref{co6}), where the 
interpretation is completely different. In addition, as shown in
(\ref{t4}), for off-shell gravitons the winding mass $w$ is actually
replaced by the (squared) momentum of the graviton, $k^2$.

\subsection{Three gauge bosons - one winding graviton amplitude}

This (tree-level) amplitude is of direct interest 
for accelerator searches and was
calculated at the effective-field theory level in \cite{GRW}. 
It is however
important to have a full string expression in order to control the
string corrections for energies close enough to the string scale $M_s$.
Moreover, this computation allows an off-shell continuation of the form
factors (\ref{t4}), (\ref{co8}) for one of the two gauge bosons.

The amplitude involves the correlation function of three gauge vertex
operators of polarisations $\e_i$ and momenta 
$p_i$ ($i=1,2,3$) and of a
massive winding-type graviton of polarisation $\e_4$ and momentum
$p_4$. 
The three conformal Killing vectors allow us to fix the positions
of the gauge vertex operators on the boundary of the disk $y_1=0 , y_2=1$ and
$y_3 = \infty$. Then the position of the graviton is unfixed on the disk,
represented as usual as the upper half plane. Some details about this
computation in type I superstring are displayed in the Appendix C.
We choose to use the 0-picture vertex operators for the gauge
bosons and the (-2) picture vertex on the disk (corresponding to the
(-1,-1) picture on the sphere) for the graviton vertex operator
\cite{fms}. Therefore we have
\begin{equation}
V_i^{\mu_i}= g_s^{1/2} \lambda^{a_i} 
:(i\partial X^{\mu_i} (y_i)+{2p_i \over M_s^2} \psi
\psi^{\mu_i}(y_i)) e^{2ip_i X(y_i)}: \ ,
\end{equation}
for the gauge bosons and 
\be
V^{\mu\nu}\!=\!V^\mu(z)V^\nu(\bar z)\!=\! g_s
:e^{-\phi(z)}\psi^\mu(z)e^{ip_4X(z)+iw Y(z)}:
:e^{-\phi(\bar z)}\psi^\nu(\bar z)e^{ip_4 X(\bar z)-iw Y(\bar z)}:
 \ee
for the graviton, where $\psi^{\mu}$ and $\phi$ 
are the world-sheet fermions and the bosonised ghosts, respectively. 
The amplitude of interest reads
\be
A_4\!=\! g_s^{-1} N_4 tr(\lambda_{a_1}\lambda_{a_2}\lambda_{a_3})\int_{C^+}d^2z<c(y_1)c(y_2)c(y_3)>
<\prod_{i=1}^{3}\e_iV_i(y_i)\e_4 V(z)\e_4 V(\bar z)>
+{1\leftrightarrow 2} \ , \label{amplit}
\ee
where we introduced the normalization constant, to be fixed by unitarity.
By using the Mandelstam variables (\ref{ma}), the kinematics of the
amplitude is summarized by the equations
\ba
s &=& -2p_1.p_2 = -2p_3p_4+w^2 \ , \ t=-2p_2.p_3=-2p_1.p_4+w^2 \ , \nonumber\\
u &=& -2p_1.p_3=-2p_2.p_4+w^2 \ , \ s+t+u=w^2 \ .
\ea
The details of the calculation are rather long and some of the steps are
sketched in Appendix C\footnote{ We added also in Appendix C the similar
but much simpler computation of an amplitude in the bosonic string of
three open string tachyons and open winding closed string tachyon, which
has similar analyticity properties to the supersymmetric amplitude we
discuss here.}.
 The final result can be put in the form
\begin{equation}
A_4\!=\!{g_{YM} \over\sqrt{\pi} M_P} \ 2^{-{w^2 \over M_s^2}} \ 
tr([\lambda^{a_1},\lambda^{a_2}]
\lambda^{a_3}) K { \Gamma\left(-{w^2 \over 2M_s^2} +{1 \over 2} \right)
\Gamma\left({-s \over 2M_s^2} \right)
\Gamma\left({-t \over 2M_s^2} \right)
\Gamma\left({-u \over 2M_s^2} \right) \over
\Gamma\left({s-w^2 \over 2M_s^2}+1 \right)
\Gamma\left({t-w^2 \over 2M_s^2}+1\right)
\Gamma\left({u-w^2 \over 2M_s^2}+1\right)} \ ,\label{resa}
\end{equation}
where $K$ is a kinematical factor displayed in  Appendix C and $N_4$ was
determined by unitarity from the three gauge bosons amplitude and two
gauge bosons -- one graviton amplitude. In order to
make connection with the field-theory result, we notice that we can
actually rewrite (\ref{resa}) in the form
\be
A_4\!=\!{1 \over \sqrt{\pi}} \ 2^{-{w^2 \over M_s^2}} \ 
{ \Gamma\left(-{w^2 \over 2M_s^2} +{1 \over 2} \right)
\Gamma\left({-s \over 2M_s^2}+1 \right)
\Gamma\left({-t \over 2M_s^2}+1 \right)
\Gamma\left({-u \over 2M_s^2}+1 \right) \over
\Gamma\left({s-w^2 \over 2M_s^2}+1 \right)
\Gamma\left({t-w^2 \over 2M_s^2}+1\right)
\Gamma\left({u-w^2 \over 2M_s^2}+1\right)} \ A_4^{FT} \ , \label{resa1}  
\ee
where $A_4^{FT}$ turns out to be exactly the field-theory amplitude
\cite{GRW}. The full string result (\ref{resa1}) obviously reduces to
the field theory result in the low energy $s,t,u << M_s^2$ and low 
graviton mass $w^2 << M_s^2$ limit. The analytic structure of the string
amplitude shows the presence of poles for $s,t,u=(2n-2) M_s^2$ for $n$ a
positive integer, corresponding to tree-level open modes
exchanges in the $s$, $t$ and $u$ channel, respectively. 
Moreover, we find, like in the previous Section, poles for graviton
winding masses
$w^2=(2n-1)M_s^2$, interpreted as a tree-level mixing between the
graviton and the gauge singlets present at the odd open string levels,
which couples afterwards to the gauge fields. We find also interesting
zeroes of the amplitudes for very heavy gravitons
$w^2=s+2n M_s^2$ or similar equations obtained by the replacement 
$s \rightarrow t,u$, giving interesting selection rules. By using
(\ref{resa1}) we are now able to extend the form factor (\ref{t4}) to
the case where one of the  gauge bosons is off-shell 
\be
g(p_1,p_2,p) = {1 \over M_P \sqrt{\pi}} \ 2^{-p^2 \over M_s^2}{\Gamma
(-p^2/2M_s^2+1/2) \over
\Gamma (-p_1p_2/2M_s^2+1)} \ , \label{form}
\ee
the result displayed in (\ref{I9}).
An important question is certainly the string deviations in (\ref{resa1})
from the field theory result $A_4^{FT}$. The energy
corresponding to the first string resonance is, in the s-channel,
$s=2M_s^2$ and similarly for $t$ and $u$, meaning that field theory 
computations certainly break down for energies above. For energies well below
this value $s,t,u,w^2 << M_s^2$, it is easy to find the corrections to 
the field-theory computation by performing a power-series expansion 
in (\ref{resa1}).
The first corrections turn out to be of the form
\be
A_4 = (1+ {\zeta(2) \over 4} {w^4 \over M_s^4}
+{\zeta(3)\over 4}{stu+w^6\over M_s^6}+
\cdots )  A_4^{FT} \ 
\ee
which, after T-duality in order to make connection with the notation in
the Introduction, becomes
\be
A_4 = (1+ {\pi^2 \over 24} {m^4 \over (R_{\perp}M_s)^4}+ \cdots )  A_4^{FT} 
\ . \label{dev}  
\ee 
This result can be interpreted as a modification of
the effective coupling of massive graviton to matter
\begin{equation}
{{1}\over{M_P}}\rightarrow{{1}\over{M_P}}(1+{\zeta(2) \over 4} 
{w^4 \over M_s^4}).
\end{equation}
Notice that the first correction to the amplitude with 
a massless graviton (of fixed energy) is of order $E^6/M_s^6$,
so the deviation from the field theoretical result 
is first expected to be seen from massive gravitons.

An experimentally more useful way to define deviations from the field
theory result is in the integrated cross-section $\sigma$, obtained by summing
over all graviton masses, up to the available energy $E$  
\be
\sigma = \sum_{m_1 \cdots m_6=0}^{R_{\perp}E} |A_4|^2 \quad , \quad 
\sigma^{FT} = \sum_{m_1 \cdots m_6=0}^{R_{\perp}E} |A_4^{FT}|^2
\ , \label{sigma}
\ee 
where $\sigma^{FT}$ is the corresponding
field theory value. Surprisingly enough, terms of order $E^2$ (or $m^2$
in (\ref{dev})) ar absent in (\ref{sigma}) and therefore at low energies
the string corrections are smaller than expected, of order 
\be
{\sigma-\sigma^{FT} \over \sigma^{FT}} \sim {E^4 \over M_s^4} \ . \label{dev1}
\ee
However, as mentioned above, strong deviations appear close to the value
$E^2=2M_s^2$, where the first string resonance appear and the field
theory approach breaks down.
  
\appendix

\section{Jacobi functions and their properties}

For the reader's convenience we collect in this Appendix the
definitions, transformation properties and some identities among the
modular functions that are used in the text. The Dedekind function is
defined by the usual product formula (with $q=e^{2\pi i\tau}$)
\be
\eta(\tau) = q^{1\over 24} \prod_{n=1}^\infty (1-q^n)\ ,
\ee
whereas the Jacobi $\vartheta$-functions with general characteristic and
arguments  are
\be
\vartheta [{\a \atop \b }] (z,\tau) = \sum_{n\in Z}
e^{i\pi\tau(n-\a)^2} e^{2\pi i(z- \b)(n-\a)}\ .
\ee
We give also the product formulae for the four special $\vartheta$-functions
\be\eqalign{
\ \ \vartheta_1(z,\tau) &\equiv \vartheta \left[{{1\over 2} \atop {1\over 2} }\right]
  (z,\tau) = 2q^{1/8}{\rm sin}\pi z\prod_{n=1}^\infty
  (1-q^n)(1-q^ne^{2\pi i z})(1-q^ne^{-2\pi i z}) \ , \cr
\ \ \vartheta_2(z,\tau) &\equiv \vartheta \left[{{1\over 2} \atop 0 }\right]
  (z,\tau) = 2q^{1/8}{\rm cos}\pi z\prod_{n=1}^\infty
  (1-q^n)(1+q^ne^{2\pi i z})(1+q^ne^{-2\pi i z}) \ , \cr
\ \ \vartheta_3(z,\tau) &\equiv \vartheta \left[{0 \atop 0 }\right]
  (z,\tau) = \prod_{n=1}^\infty
  (1-q^n)(1+q^{n-1/2}e^{2\pi i z})(1+q^{n-1/2}e^{-2\pi i z}) \ , \cr
\ \ \vartheta_4(z,\tau) &\equiv \vartheta \left[{0 \atop {1\over 2} }\right]
  (z,\tau) = \prod_{n=1}^\infty
  (1-q^n)(1-q^{n-1/2}e^{2\pi i z})(1-q^{n-1/2}e^{-2\pi i z}) \ .
}
\ee
The modular properties of these functions are described by
\be
\eta(\tau+1) = e^{i\pi/12}\eta(\tau)\ \ , \ \
\vartheta \left[{\a \atop {\b}}\right] \left({z} ,
  {\tau+1}\right)=
e^{-i\pi\a(\a-1)}\vartheta 
\left[{\a \atop {\a+\b-{1\over 2}}}\right] \left({z} \ ,
  {\tau}\right) \nonumber 
\ee
\be
\eta(-1/\tau) = \sqrt{-i\tau}\; \eta(\tau)\ \ , \ \ 
\vartheta \left[{\a \atop {\b}}\right] \left({z \over \tau}, {-1 \over \tau}\right)=
\sqrt{-i \tau} \ e^{2 i \pi \a \b +{i \pi z^2 / \tau}} \ 
\vartheta \left[{{\b} \atop - \a}\right] (z, \tau ) \ . \label{f8}
\ee
\section{Box diagrams in 4D}

We consider the box amplitude $B(s,t,m_1,\dots,m_4)$
in four dimensions (see Figure 7) in the euclidean formulation
\begin{equation}
{{M_s^{-4}\pi^42}\over{3}}
\int d\tau_2 \tau_2 d^{4}\eta \delta (1-\sum_i \eta_i)
e^{{2\pi\tau_2 \over M_s^2} (s\eta_1\eta_3+t\eta_2\eta_4)}
e^{-{2\pi\tau_2 \over M_s^2} \sum_i m_i^2\eta_i}\equiv
{{\pi^2}\over{6}}B' \ .
\end{equation}
\begin{figure}
\vspace{5 cm}
\special{hscale=60 vscale=60 voffset=0 hoffset=120
psfile=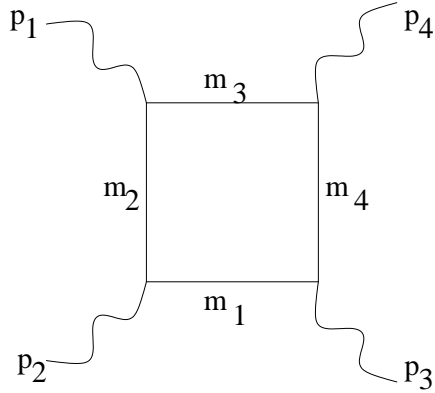}
\caption{The box diagram with particles of masses $m_i$ in the loop}
\end{figure}

It is possible to perform the integral over $\tau_2$ as well
as the integration over two $\eta$ variables
and obtain
\ba
\!\!\!&&\!\!\!B'\!=\!\int_0^1d\eta_1\int_0^{1-\eta_1}\! d\eta_2 \nonumber\\
&&\!\!\!\!\!{{1-\eta_1-\eta_2}\over
{[(1\!-\!\eta_1\!-\!\eta_2)(m_4^2\!-\!t\eta_2)\!+\!\eta_1m_1^2\!+\!\eta_2m_2^2]
[(1\!-\!\eta_1\!-\!\eta_2)(m_3^2\!-\!s\eta_1)\!+\! \eta_1m_1^2\!+\!
\eta_2m_2^2]}} \ .
\end{eqnarray}
For equal masses the expression simplifies to
\begin{equation}
B'=\int_0^1d\eta_1\int_0^{1-\eta_1}d\eta_2 
{{1-\eta_1-\eta_2}\over
{[m^2-t\eta_2(1-\eta_1-\eta_2)][m^2-s\eta_1(1-\eta_1-\eta_2)]}} \ .
\end{equation}
If $-s<<m^2$ and $-t<<m^2$ we get
the dominant contribution $B'=1 /{6m^4}+\dots$.
In the opposite limit when $m^2<<-s$ and $m^2<<-t$,
it is convenient to change variables to
$\alpha=\eta_2(1-\eta_1-\eta_2)$ and
$\beta=\eta_1(1-\eta_1-\eta_2)$, so that the integral becomes
\begin{equation}
B'=2\int_0^{1/4}d\alpha\int_0^{1/4-\alpha}d\beta
{{1}\over{\sqrt{1-4(\alpha+\beta)}}}\ 
{{1}\over{m^2-t\alpha}} \ {{1}\over{m^2-s\beta}} \quad .
\end{equation}
In the limit where $m$ is very small one can neglect the first
factor in the integral and we get
\begin{equation}
B'={{1}\over{st}}\ln{{-s}\over{4m^2}}\ln{{-t}\over{4m^2}}+\dots.
\end{equation}
Another case encountered is the one where 
$m_4=M$ is large and the other masses are equal ($\mu$) and small.
The integral is approximated by
\begin{equation}
{{1}\over{M^2}}
\int_0^1d\eta_1\int_0^{1-\eta_1}d\eta_2 {{1}\over
{\mu^2-s\eta_1\eta_2}}={{-1}\over{M^2s}}\int_0^1
{d \eta \over \eta } \ln[1-{{s\eta(1-\eta)}\over{\mu^2}}] \ ,
\end{equation}
which is approximately equal to
\begin{equation}
{{-1}\over{2sM^2}}\ln^2{{-s}\over{\mu^2}}.\label{b7}
\end{equation}
The case where $m_2$ is very large 
and the other masses small gives the same answer and the
one where $m_1$ or $m_3$ are very large is obtained by changing $s$
into $t$.
 Another case of interest corresponds to $m_4=m_3=M$ and 
 $m_1=m_2=\mu$. The box diagram to leading order is given in this case by
 \begin{equation}
 B'={{1}\over{M^4}}\int_0^1d\eta_1\int_0^{1-\eta_1}d\eta_2
{{1}\over{1-(\eta_1+\eta_2)(1+\mu^2/M^2)}}=
{{1}\over{M^4}}\ln {{M^2}\over{\mu^2}} \ .
\end{equation}
We are now in a position to calculate the leading correction,
$\Delta A_1$,
to the amplitude $A_1$  discussed in Section 3.5.
 It is the sum of the three terms 
\begin{eqnarray}
A^{(1,1)}_1(l_0)(s,t)&=&{8 g_{YM}^4 \over M_s^4}
\int_{1/l_0}^{\infty}d\tau_2 \tau_2 \int_{\eta_i>0} 
d^{4}\eta
\delta(1-\sum_i\eta_i)
e^{{2\pi \tau_2 \over M_s^2} (s\eta_1\eta_3+t\eta_2\eta_4)}
e^{-{2\pi\tau_2 \mu^2 \over M_s^2}} \nonumber\\
&& \{-(s/M_s^2)(\ln f(\eta_2)+\ln f(\eta_4))
-(t/M_s^2) (\ln f^T(\eta_1)+\ln f^T(\eta_3))\} \ , \nonumber \\
A^{(2,1)}_1(l_0)(u,t)&=&{8 g_{YM}^4 \over M_s^4}
\int_{1/l_0}^{\infty}d\tau_2 \tau_2 \int_{\eta_i>0} 
d^{4}\eta
\delta(1-\sum_i\eta_i)
e^{{2\pi \tau_2 \over M_s^2} (u\eta_1\eta_3+t\eta_2\eta_4)}
e^{-{2\pi\tau_2 \mu^2 \over M_s^2}} \nonumber\\
&& \{-(u/M_s^2) (\ln f^T(\eta_2)+\ln f^T(\eta_4))
-(t/M_s^2) (\ln f^T(\eta_1)+\ln f^T(\eta_3))\} \ , \nonumber \\
A^{(3,1)}_1(l_0)(u,s)&=&{8 g_{YM}^4 \over M_s^4}
\int_{1/l_0}^{\infty}d\tau_2 \tau_2 \int_{\eta_i>0} 
d^{4}\eta \delta(1-\sum_i\eta_i)
e^{{2\pi \tau_2 \over M_s^2} (u\eta_1\eta_3+s\eta_2\eta_4)}
e^{-{2\pi\tau_2 \mu^2 \over M_s^2}} \nonumber\\
&& \{-(u/M_s^2) (\ln f^T(\eta_2)+\ln f^T(\eta_4))
-(s/M_s^2) (\ln f(\eta_1)+\ln f(\eta_3))\} \ .\label{cor}
\end{eqnarray}
Recall that 
\ba
f(\eta)=(1-e^{-2\pi\eta\tau_2})
\prod_{n=1}^{\infty}(1-e^{-2\pi(n-\eta)\tau_2})
(1-e^{-2\pi(n+\eta)\tau_2}) \ , \\
f^T(\eta)=(1+e^{-2\pi\eta\tau_2})
\prod_{n=1}^{\infty}(1+e^{-2\pi(n-\eta)\tau_2})
(1+e^{-2\pi(n+\eta)\tau_2}) \ ,
\ea
The developpement of $\ln f(\eta)$ as
\begin{equation}
\ln f(\eta)=-\sum_{m=1}^{\infty}
{{e^{-2\pi m\tau_2\eta_2}}\over{m}}+\dots
\end{equation}
and a similar expansion of $\ln f^T$
shows that every term in the above sum represents a box diagram with
a massive string mode in one leg and the other three particles
of mass $\mu$, while the dots represent box diagrams
with more than one leg having a massive string mode and are thus
of higher order.
As explained in section 3.5,
 since the diagrams are UV convergent,
in order to get the $l_0$ independent terms it suffices to calculate 
the corresponding box diagram.
 From (\ref{b7}) we obtain the leading contribution to 
$A_1^{(i,1)}$ as
\ba
A_1^{(1,1)}(s,t)&=&-{{g_{YM}^4}\over{3M_s^4}}\left(
\ln^2{{-s}\over{\mu^2}}-{{1}\over{2}}
\ln^2{{-t}\over{\mu^2}}\right),\
A_1^{(2,1)}(u,t)={{g_{YM}^4}\over{3M_s^4}}\left(
{{1}\over{2}}\ln^2{{-u}\over{\mu^2}}+{{1}\over{2}}
\ln^2{{-t}\over{\mu^2}}\right),\nonumber\\
A_1^{(3,1)}(u,s)&=&-{{g_{YM}^4}\over{3M_s^4}}\left(
\ln^2{{-s}\over{\mu^2}}-{{1}\over{2}}\ln^2{{-u}\over{\mu^2}}\right),
\label{b10}
\ea
where we have used $\sum 1/m^2=\pi^2/6$.
The sum of the three terms in (\ref{b10}) gives
\be
\Delta A_1={{g_{YM}^4}\over{3M_s^2}}\left(
\ln{{s}\over{t}}\ln{{st}\over{\mu^2}}+
\ln{{s}\over{u}}\ln{{su}\over{\mu^2}}\right),
\ee
which is the result used in the text (\ref{f4}).
Notice that the terms in $\ln^2\mu^2$ have cancelled in $\Delta
A_1$.


\section{The type I disk amplitude}

In order to compute the amplitude (\ref{amplit})
depicted in Figure 8, it is convenient first to
write the vertex operators with the aid of Grassmann variables $\theta_i$ and $\phi_i$ as 
\begin{eqnarray}
\e_iV_i&\!\!=\!\!&\int d\theta_id\phi_i
e^{2ip_iX(y_i)+i\theta_i\phi_i\e_i.\partial X_i-
{2 \over M_s} \theta_ip_i\psi_i+{1 \over M_s} \phi_i\e_i.\psi} \ ,
\\
\e_4V_4(z)&\!\!\!=\!\!\!&\int d\phi_4 e^{ip_4.X(z)+iw.Y(z)
+\phi_4\e_4.\psi(z)} \ , \
\e_4V_4(\bar z)\!\!=\!\!\int d\bar\phi_4 e^{ip_4.X(\bar z)-iw.Y(\bar z)
+\bar \phi_4\e_4.\psi(z)} \ . \nonumber 
\end{eqnarray}
The correlation functions are then easily calculated with the aid
of
\begin{eqnarray}
<\psi^\mu(z_1)\psi^\nu(z_2)>&\!\!=\!\!&\eta^{\mu\nu}{{1}\over{z_1-z_2}} \ , \\
<c(y_1)c(y_2)c(y_3)>&\!\!\!=\!\!\!&(y_1-y_2)(y_1-y_3)(y_2-y_3) \ , \quad
<e^{-\phi(z)}e^{-\phi(\bar z)}>\!=\!{{1}\over{z-\bar z}} \ . \nonumber
\end{eqnarray}
Then we fix the positions $y_i$ to $0,1$ and $\infty$
and calculate the integral over the Grassmann variables. The
resulting amplitude involve integrals of the form
\begin{equation}
I_n(\alpha,\beta,\gamma)=\int_{-\infty}^{\infty}dx
\int_{0}^{\infty}dy \ x^n y^\alpha (x^2+y^2)^\beta
\left[(1-x)^2+y^2\right]^\gamma \ ,
\end{equation}
where $n=0,1$ and $\alpha$, $\beta$ and $\gamma$ are
real. These integrals can be calculated using the standard tricks
to yield
\ba
I_0(\alpha,\beta,\gamma)&\!\!=\!\!&
{{\sqrt{\pi}}\over{2}}{{\Gamma\left({{\alpha+1}\over{2}}\right)}
\over{\Gamma(-\beta)\Gamma(-\gamma)}}
\Gamma\left(\!-\! \beta \!-\! \gamma \!-\! {{\alpha}\over{2}}\!-\! 1\right)
B\left(\gamma \!+\! {{\alpha}\over{2}}\!+\!1,\beta \!+\! {{\alpha}\over{2}}\!+\!1
\right) \ , \nonumber \\
I_1(\alpha,\beta,\gamma)&\!\!\!=\!\!\!&
{{\sqrt{\pi}}\over{2}}{{\Gamma\left({{\alpha+1}\over{2}}\right)}
\over{\Gamma(-\beta)\Gamma(-\gamma)}}
\Gamma\left(\!-\! \beta \!-\! \gamma \!-{{\alpha}\over{2}} \!-\! 1\right)
B\left(\gamma \!+\! {{\alpha}\over{2}}\!+\! 1,\beta \!+\!
{{\alpha}\over{2}} \!+\!2
\right) \ .
\ea
With the aid of theses integrals the amplitude can be calculated
and after some arrangements and summing
the two cyclically inequivalent permutations of the open states
it can be put in the form (\ref{resa})
with the kinematical factor $K$
given by
\begin{eqnarray}
K&=&
s(u+t)
\Big(
\e_1.\e_3\e_2.p_3\e_4.p_1\e_4.p_2-
\e_1.p_3\e_2.\e_3\e_4.p_1\e_4.p_2 -
\e_1.\e_4\e_2.\e_4\e_3.p_2p_1.p_3 \nonumber \\
&+& \e_1.\e_4\e_2.\e_3\e_4.p_2p_1.p_3
\e_1.\e_4\e_2.\e_4\e_3.p_1p_2.p_3-
\e_1.\e_3\e_2.\e_4\e_4.p_1p_2.p_3\nonumber\\
&-&
\e_1.\e_4\e_2.p_3\e_3.p_1\e_4.p_2+
\e_1.p_3\e_2.\e_4\e_3.p_2\e_4.p_1\Big)\nonumber\\
&+&{{1}\over{2}}(st \e_3.p_1-su\e_3.p_2)
\Big( \e_1.\e_2.\e_4.p_1\e_4.p_2-
\e_1.\e_4\e_2.\e_4p_1.p_2
\e_1.\e_4\e_2.p_1\e_4.p_2 \nonumber \\
&-& \e_1.p_2\e_2.\e_4\e_4.p_1\Big) +(1,2,3,4)\rightarrow (3,1,2,4)+
(1,2,3,4) \rightarrow(2,3,1,4).
\end{eqnarray}
\begin{figure}
\vspace{4 cm}
\special{hscale=60 vscale=60 voffset=0 hoffset=120
psfile=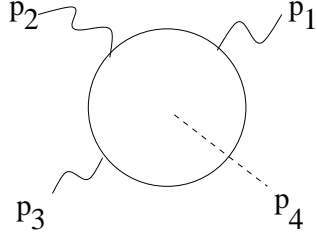}
\caption{The disk amplitude with three open string particles and one
closed string particle.}
\end{figure}
For pedagogical reasons, we present here also a similar but much simpler 
computation in the bosonic string, the four-point function of three open
string tachyons and one winding state closed string tachyon. The vertex
operators $V_i$ for open tachyons and $V_4$ for the winding state closed
tachyon in this case are 
\ba
V_i &=& g_s^{1/2} \lambda^{a_i} :e^{2p_i X(y_i)}: \ , \nonumber \\ 
V_4 &=& g_s :e^{ip_3X(z)+iwY(z)}: 
:e^{ip_3 X({\bar z})-iwY({\bar z})}: \ . 
\ea
By using unitarity arguments as in Section 4 in order to fix the
overall normalization constant, the amplitude to compute becomes therefore
\be
A= {g_{YM} \over \pi M_P} tr(\lambda_{a_1}\lambda_{a_2}\lambda_{a_3})
\int_{C^+} d^2 z |z|^{2p_1p_4 \over M_s^2}
|1-z|^{2p_2p_4 \over M_s^2} |z-{\bar z}|^{{p_4^2 \over 2M_s^2}-
{w^2 \over 2M_s^2}+2}+1\leftrightarrow 2 \ , 
\ee
where the complex integral is in the upper half complex plane. 
The kinematics of the process is described by
\ba
&&p_1^2=p_2^2=p_3^2=M_s^2 \ , \ p_4^2+w^2= 4M_s^2 \ , \nonumber \\
&&s=M_s^2-w^2+2 (p_1+p_2)p_4 \ , \ t = -5M_s^2 + w^2-2p_2p_4 \ ,
\nonumber \\
&&u=  -5M_s^2 + w^2-2p_1p_4 \ , \ s+t+u = -5 M_s^2 +w^2 \ . 
\ea
The simplest way to compute the amplitude is to use equalities of the type
\be
({1 \over z {\bar z}})^a= {1 \over \Gamma (a)} \int_0^{\infty} dt
t^{a-1} e^{-t z} \ , 
\ee
which was also used in order to obtain (C.4) from (C.3).
The final result is
\begin{equation}
A \!\!=\!\! {g_{YM} \over{\sqrt{\pi} M_P}} \ 2^{\!-{w^2 \over M_s^2}\!+\!1}
 tr(\{\lambda^{a_1}\! ,\lambda^{a_2}\}
\lambda^{a_3}) { \Gamma\left(\!-\!{w^2 \over 2M_s^2}\!+\! {3 \over 2} \right)
\Gamma\left({-s \over 2M_s^2}\!-\!{1 \over 2} \right)
\Gamma\left({-t \over 2M_s^2}\!-\!{1 \over 2} \right)
\Gamma\left({-u \over 2M_s^2}\!-\!{1 \over 2} \right) \over
\Gamma\left({s-w^2 \over 2M_s^2}\!+\!{5 \over 2} \right)
\Gamma\left({t-w^2 \over 2M_s^2}\!+\!{5 \over 2}\right)
\Gamma\left({u-w^2 \over 2M_s^2}\!+\!{5 \over 2} \right)} \ .
\end{equation}

\vfill
\eject




\newpage
\begin{thebibliography}{99}

\bibitem{gsw} M.B. Green, J.H. Schwarz and E. Witten, {\it Superstring
Theory}, Vol. I,II, Cambridge University Press, 1987.
 
\bibitem{polchinski} J. Polchinski, {\it String Theory}, Vol. I,II, 
Cambridge University Press, 1998.

\bibitem{antoniadis} I. Antoniadis, \PLB{246}{90}{377}; I. Antoniadis
and K. Benakli, \PLB{326}{94}{69};  I. Antoniadis, K. Benakli and M. Quiros, 
\PLB{331}{94}{313}. 

\bibitem{witten} E. Witten, \NPB{471}{96}{135}, J.D. Lykken, \PRD{54}{96}{3693}.

\bibitem{ADD} N. Arkani-Hamed, S. Dimopoulos and G. Dvali, \PLB{429}{98}{263},
hep-ph/9807344.

\bibitem{DDG} K.R. Dienes, E. Dudas and T. Gherghetta, \PLB{436}{98}55,
\NPB{537}{99}{47}, hep-ph/9807522; C. Bachas, hep-th/9807415; 
D. Ghilencea and G.G. Ross, \PLB{442}{98}{165}; Z. Kakushadze, hep-th/9811193; 
A. Delgado and M. Quir{\'o}s, hep-ph/9903400;
Z. Kakushadze and T.R. Taylor, hep-th/9905137;
I. Antoniadis, C. Bachas and E. Dudas, hep-th/9906039.

\bibitem{AADD}I. Antoniadis, N. Arkani-Hamed, S. Dimopoulos and G. Dvali,
\PLB{436}{98}{263}, G. Shiu and S.-H.H. Tye, \PRD{58}{98}{106007};
K. Benakli, \PRD{60}{99}{104002}; 
C. Burgess, L.E. Ib{\'a}{\~n}ez and F. Quevedo, hep-ph/9810535;
I. Antoniadis and C. Bachas, hep-th/9812093;
I. Antoniadis and B. Pioline, \NPB{550}{99}{41}.

\bibitem{GRW} G.F. Giudice, R. Rattazzi and 
J.D. Wells, \NPB{544}{99}{3}; E.A. Mirabelli, M. Perelstein and
M.E. Peskin, \PRL{82}{99}{2236}; T. Han, J.D. Lykken and R.J. Zhang,
\PRD{59}{99}{105006}; J.L. Hewett, \PRL{82}{99}{4765}; T.G. Rizzo, 
\PRD{59}{99}{115010}, \PRD{60}{99}{075001}; P. Mathews, S. Raychaudhuri
and K. Sridhar, \PLB{450}{99}{343} and hep-ph/9904232; M. Besancon, hep-ph/9909364. 

\bibitem{sagnotti} A. Sagnotti, hep-th/9302099; C. Angelantonj,
M. Bianchi, G. Pradisi, A. Sagnotti
and Ya.S. Stanev, \PLB{387}{96}{743}; G. Zwart, \NPB{526}{98}{378};
Z. Kakushadze, G. Shiu and S.H.H. Tye, \NPB{533}{98}{25};
G. Aldazabal, A. Font, L.E. Iba\~nez and G. Violero, \NPB{536}{98}{29}.

\bibitem{dua} E. Cremmer and J. Scherk, \NPB{50}{72}{222};
L. Clavelli and J.A. Shapiro, \NPB{57}{73}{490};
C.G. Callan, C. Lovelace, C.R. Nappi and S.A. Yost, 
\NPB{293}{87}{83}, \NPB{308}{88}{221};
J. Polchinski and Y. Cai, \NPB{296}{88}{91}.

\bibitem{HK} A. Hashimoto and I.R. Klebanov, \PLB{381}{96}{437},
{\it Nucl. Phys. Proc. Suppl.} {\bf B55} {1997} 118.

\bibitem{gm} M.R. Garousi and R.C. Myers,  \NPB{475}{96}{193}. 

\bibitem{gsb}M.B. Green, J.H. Schwarz and L. Brink, \NPB{198}{82}{474}.

\bibitem{gs}M.B. Green and J.H. Schwarz, \NPB{198}{82}{441}.

\bibitem{dhp} E. D'Hoker and D.H. Phong, \NPB{440}{95}{24}. 

\bibitem{ts} A.A Tseytlin, \PLB{367}{96}{84},
\NPB{467}{96}{383}; C. Bachas and E. Kiritsis,  
{\it Nucl. Phys. Proc. Suppl.} {\bf B55} (1997) 194.

\bibitem{bkny} M. Bando, T. Kugo, T. Nagoshi and K. Yoshioka,
hep-ph/9906549; J. Hisano and N. Okada, hep-ph/9909555.

\bibitem{fms} D. Friedan, E. Martinec and S. Shenker, \NPB{271}{86}{93}.

\end{thebibliography}
\end{document}